\documentclass[aps,floatfix,onecolumn,a4paper,noshowpacs, nofootinbib,superscriptaddress,12pt]{revtex4}

\usepackage{booktabs}
\usepackage{multirow}
\usepackage{multirow,array}
\usepackage[utf8]{inputenc}
\usepackage{amsmath,amssymb,graphicx,bm}
\usepackage{geometry}

\usepackage{siunitx}
\usepackage{hyperref}
\hypersetup{colorlinks=true, linkcolor=blue, citecolor=blue, urlcolor=blue}
\usepackage{graphicx,float,tikz}
\usepackage[all]{xy}
\usepackage{bm,amsmath,upgreek,bigints}
\newcommand{\beq}{\begin{eqnarray}}
\newcommand{\eeq}{\end{eqnarray}}

\usepackage{amssymb}
\usepackage{color}
\usepackage{epsfig,bm}		
\usepackage{graphicx,epstopdf}
\usepackage{subfigure,upgreek}
\usepackage{pdfpages}
\usepackage{multirow}
\newcommand{\RR}{\mathbb{R}}
\newcommand{\sie}{Shannon information entropy}

\newcommand{\blt}{\textcolor{black}}
\newcommand{\bltt}{\textcolor{black}}
\usepackage{bookmark,textgreek}
\usepackage{hyperref,color,xcolor}
\definecolor{lightgray}{gray}{0.95}
\hypersetup{hidelinks,colorlinks=true,breaklinks=true,urlcolor= blue}
\hypersetup{%
    colorlinks = true,
    linkcolor  = blue,
    citecolor = cyan,
  }
\usepackage{siunitx}
\definecolor{headergray}{gray}{0.90}

\usepackage[font=small,labelfont=bf,labelsep=space]{caption}
\usepackage{natbib}
\setcitestyle{square,numbers}

\newcommand\orcidroldao{{\href{https://orcid.org/0000-0003-3978-532X}{\orcidicon}}}
\newcommand\orcidrafael{{\href{https://orcid.org/0000-0003-0960-5236}{\orcidicon}}}

\newcommand{\orcidicon}{%
	\begin{tikzpicture}
	\draw[lime, fill=lime] (0,0)
		circle [radius=0.16]
		node[white] {{\fontfamily{qag}\selectfont \tiny ID}};
	\draw[white, fill=white] (-0.0625,0.095)
		circle [radius=0.007];
	\end{tikzpicture}	\hspace{-2mm}
}

\begin{document}

\title{Digit anomalies in the hadronic mass spectrum, {}{classical and quantum information entropies}, and the dynamical QCD scale}
\author{R. da Rocha\orcidroldao\!\!}
\email{roldao.rocha@ufabc.edu.br}
\affiliation{Federal University of ABC, Center of Mathematics, Santo Andr\'e, S\~ao Paulo 09580-210, Brazil}
\author{R. D. Vilela\orcidrafael\!\!}
\email{rafael.vilela@ufabc.edu.br}
\affiliation{Federal University of ABC, Center of Mathematics, Santo Andr\'e, S\~ao Paulo 09580-210, Brazil}
\begin{abstract}
Quantum Chromodynamics (QCD) has an emergent dynamical energy scale $\Lambda_{\rm QCD}$ 
which sets the threshold between perturbative and nonperturbative regimes. This
characteristic scale causes hadronic masses to cluster within certain mass ranges,
instead of following a uniform distribution. Analyzing the Shannon information
entropy underlying the hadronic mass spectrum, {}{and also other classical information entropies}, provides novel insight into this phenomenon, revealing a pronounced deviation from the law of anomalous numbers.
This deviation quantifies the emergence of the dynamical scale in strongly interacting systems, also encoding the information-entropy cost associated with the breaking
of scale invariance in QCD. {}{Quantum entanglement entropy also reveals correlations across energy scales through the Shannon entropy of the bipartitioned probability distribution, reflecting how the emergence 
of $\Lambda_{\textsc{QCD}}$ shapes the hierarchical structure of the hadronic mass spectrum.}
\end{abstract}

\maketitle
\section{Introduction}
\label{intro}
Shannon’s seminal paper \cite{shannon1948} forged information into a quantitative concept that underlies communication theory, establishing the foundation for error-correcting codes and data-compression algorithms. Shannon information entropy categorizes the evolution of dynamical systems among regular, chaotic, and completely random regimes, making it particularly useful in nuclear reactions \cite{Karapetyan:2018oye,Karapetyan:2023kzs,Ma:2018wtw}. The production of hadrons in heavy-ion collisions is difficult to analyze directly, owing to the intricate and rapidly evolving dynamics of nuclear matter. Applications of \sie\ include the analysis of chaotic patterns in hadron branching processes and in the quark-gluon plasma (QGP), a short-lived, thermalized QCD phase of deconfined, ultra-dense hot nuclear matter. \sie~comprises the fundamental quantity in configurational information measures (CIMs) \cite{Gleiser:2018kbq,Gleiser:2018jpd,Bernardini:2016hvx,Witten:2018zva}, which have been explored to address physical aspects of hadronic matter in QCD. The hadronic mass spectrum have been comprehensively investigated using CIMs to estimate the mass spectrum of heavier resonances in several meson and baryon families \cite{Bernardini:2018uuy,Karapetyan:2018yhm,MartinContreras:2020cyg,Braga:2022yfe,daRocha:2024sjn,Ferreira:2019inu,Colangelo:2018mrt,Ma:2015lpa,MartinContreras:2023oqs,Ferreira:2020iry,MartinContreras:2022lxl,Guo:2024nrf}, supporting experimental data reported in the Particle Data Group (PDG) \cite{pdg}. One can also delve into several relevant aspects and applications of quantum field theory using \sie~\cite{Correa:2015lla,Gleiser:2012tu,Bazeia:2021stz,Mvondo-She:2023xel,Casadio:2022pla}.

The distribution of the full widths of hadrons was shown to satisfy the law of anomalous numbers \cite{Shao:2009zze}, also known as Newcomb--Benford’s law \cite{Benford:1938ano}, which was also employed to scrutinize nuclear structure physics \cite{Jiang:2011zza}. This law asserts that the leading digits of certain naturally occurring datasets follow a logarithmic distribution, with  smaller digits appearing more frequently. Similar patterns arise in analyzing pulsars~\cite{Shao:2010mm}, gamma-ray bursts~\cite{Lai:2024dlc}, scaling phenomena~\cite{Bera:2018smp}, and fermionic/bosonic distributions in statistical mechanics~\cite{Shao:2010rq}.

The main objective of this work is to apply the Shannon information entropy to the distribution of the hadronic mass spectrum in QCD and to show that the leading digit in hadron masses violates Newcomb--Benford’s law, which maximizes \sie~if scale invariance is assumed. The departure from the Newcomb--Benford’s law \bltt{is consistent with} the existence of a fundamental scale, $\Lambda_{\text{QCD}}$, that breaks scale invariance. It selects preferred hadronic mass ranges, thereby creating hadron mass clusters, ranging from 139~MeV to 11.1~GeV. This feature is supported by the Hagedorn pattern of the density of states \cite{Hagedorn1965}. In Appendix \ref{a12}, the Hagedorn growth, associated with QCD ultraviolet (UV) spectrum, and the mass gap intrinsic to infrared (IR) confinement, will be presented as dual manifestations of the existence of $\Lambda_{\text{QCD}}$. The leading-digit distribution of the hadronic mass spectrum will be shown in this work to have lower \sie, as compared to the Newcomb--Benford probability distribution. This offset \bltt{quantifies a statistical deviation from} scale invariance: while the Newcomb--Benford distribution corresponds to the maximal \sie\ of a scale-invariant system, the hadronic mass spectrum displays a \sie~deficit that \bltt{is compatible with} how QCD dynamics impose a preferred energy scale on the hadronic mass spectrum landscape.

{}{Beyond \sie, other classical and \bltt{information-theoretic} measures such as Kullback--Leibler and Jensen--Shannon divergences, conditional entropies, and mutual information provide quantitative measures of these deviations, revealing that baryons, in particular, carry the strongest imprint of confinement dynamics. 
The hadronic spectrum exhibits correlations across energy scales that can be captured through \bltt{information-theoretic bipartite constructions} \cite{Casini:2011kv}. By partitioning the spectrum into IR and UV domains, one can define classical bipartite entropies that quantify how QCD dynamics concentrate states in the IR and suppress correlations between the hadronic sectors containing light-flavor quarks and the ones containing at least one heavy-flavor quark. These spectral information entropies, including the Shannon entropy of the bipartitioned probability distribution, also \bltt{indicate departures from scale invariance compatible with} the emergence of $\Lambda_{\mathrm{QCD}}$.}

This paper is organized as follows: Section \ref{af1} introduces the fundamental aspects of \sie~and shows that, under the assumption of scale invariance, the Newcomb--Benford probability distribution maximizes it. Section \ref{hip} analyzes \sie~associated with the leading-digit distribution of the hadronic mass spectrum as listed in PDG \cite{pdg}, showing the breakdown of Newcomb--Benford’s law due to the scale $\Lambda_{\text{QCD}}$. This is accomplished by comparing the Shannon information entropy deficit between the observed hadron spectrum and Newcomb--Benford expected probability distribution. The analysis is split into three cases: (i) mesons, (ii) baryons, and (iii) the combined meson-baryon dataset. {}{Section \ref{sec:info} extends this analysis to additional information-theoretic measures, including Kullback--Leibler and Jensen--Shannon divergences, conditional entropies, and mutual information, which provide complementary quantifiers of the departure from scale invariance. By quantifying correlations between hadronic sector and leading-digit distributions, these measures reveal how $\Lambda_{\text{QCD}}$ \bltt{is reflected in} the statistical properties of the spectrum beyond what is captured by the Shannon entropy alone. Following this, Section \ref{sec:entanglement} introduces \bltt{bipartite information-theoretic constructions} that probe correlations across mass scales, offering an information-theoretic analogue of IR-UV correlations and shedding new light on the imprint of the QCD scale. Also, the relationship among  \bltt{bipartite spectral entropy}, \bltt{the Shannon entropy of the bipartitioned probability distribution}, hadron mass clusters, and \bltt{statistical features compatible with QCD-induced scale breaking effects} is discussed, providing an information-theoretic perspective on the emergence of $\Lambda_{\text{QCD}}$ in the hadronic mass landscape. Appendices \ref{a11} and \ref{a12} provide detailed derivations: the variational proof of Newcomb--Benford’s law as a maximum \sie~distribution; scale symmetry breaking in QCD; and an overview of hadronic mass clustering, the Hagedorn spectrum arising from counting the number of hadronic density of states, and the mass gap.
\section{Shannon information entropy under scale invariance and QCD scale invariance breaking}
\label{af1}
Information communication protocols aim to reproduce a selected message  through a channel. For an ordered  system with discrete events $H = \{E_1, E_2,\ldots, E_n\}$, respectively with probabilities $P = \{p_1, p_2,\ldots, p_n\}$, Shannon defined the information entropy of the message space as
\begin{equation}
S = -\sum_{i=1}^{\,n} p_i \ln p_i, \quad\qquad \sum_{i=1}^{n} p_i = 1,
\end{equation}
measuring the number of bits to which a message with $n$ events can be 
compressed~\cite{Witten:2018zva}. If all events are equally probable 
($p_i = 1/n$), the \sie\ reduces to $S = \ln n$, whereas selecting among a 
single possible event carries no uncertainty and yields $S = 0$. With the 
base-$e$ logarithm, \sie\ is expressed in the natural unit of entropy (nat), 
corresponding to the information associated with an event occurring with 
probability $e^{-1}$.

We aim to calculate and analyze the \sie~for the leading-digit distribution of the hadronic mass spectrum. First, under the assumption of scale invariance, 
the Newcomb--Benford distribution will be shown to be the only probability density that maximizes \sie. 
To be as general as possible, this demonstration will be implemented for the differential \sie\footnote{The demonstration of the discrete case is quite a direct consequence that follows \emph{mutatis mutandis}.}.  For a continuous probability distribution $P(x)$ defined on $\RR^+$, \sie~takes the form
\begin{equation}\label{mg1}
S[P] = - \int_0^\infty P(x) \ln P(x) \, dx.
\end{equation}
Eq. (\ref{mg1}) defines the (differential) Shannon information entropy, which can be thought of as being the continuous limit of \sie~as long as the information dimension -- measuring how the information entropy increases as the measurement resolution improves -- is controlled \cite{Gleiser:2018kbq,Gleiser:2018jpd,Bernardini:2016hvx,Witten:2018zva}. 
One can determine $P(x)$ that maximizes $S[P]$ under the assumption of scale invariance. Under scaling transformations $x \mapsto  \lambda x$, for $\lambda\in\RR^+$, scale invariance means that   $P(\lambda x)  d(\lambda x) = P(x) dx$, 
 implying that $P(x)$ is a homogeneous function of degree minus one, namely $P(\lambda x) = \frac{1}{\lambda} P(x)$, whose only nontrivial solution is given by 
\begin{equation}
P(x) = \frac{C}{x},\qquad C\in \RR. 
\end{equation}
One can also define $
y = \ln x
$, 
introducing the probability distribution in $y$-space as $Q(y) dy = P(x) dx$, in such a way that   
\begin{equation}
 \quad P(x) = \frac{Q(\ln x)}{x}.\label{pqy}
\end{equation}
Normalization $
\int_0^\infty P(x) dx = \int_{-\infty}^{\infty} Q(y) dy = 1$ is assumed. 

\sie~in logarithmic variables can be read off, as substituting Eq. (\ref{pqy}) into Eq. (\ref{mg1}) gives (introducing a cutoff for the integrand to be normalizable\footnote{Without a finite support, \sie~maximization becomes ill-defined.
The distribution $
P(x) \propto \frac{1}{x},$ with $x \in (0,\infty),$
cannot be normalized, since the integral $
\int_{0}^{\infty} \frac{dx}{x}$ diverges. 
Moreover, the left-hand side of Eq. (\ref{mg1})  
can be made arbitrarily large by spreading the probability distribution  over arbitrarily wider ranges.  Hence, the  maximization
problem demands a finite cutoff, $
x \in [x_{\textsc{min}},\,x_{\textsc{max}}],$ within which the maximum-\sie~solution consistent with
scale invariance will be given by Eq. (\ref{10}).})
\begin{align}
S 
&= - \int_{y_{\textsc{min}}}^{y_{\textsc{max}}} Q(y) \ln Q(y) \, dy + \int_{y_{\textsc{min}}}^{y_{\textsc{max}}} y Q(y) \, dy,\label{qy1}
\end{align}
with $
y_{\textsc{min}} = \ln x_{\textsc{min}},$ and $y_{\textsc{max}} = \ln x_{\textsc{max}}.$ 
The leading term on the right-hand side of Eq. (\ref{qy1}) is \sie~in log-space, measuring the spread over scales.
Scale invariance $x \mapsto \lambda x$ for the probability distribution $P(x)$ becomes translation invariance $Q(y+\ln \lambda) = Q(y)$, 
whose only normalized  solution is given by 
\begin{equation}\label{1n}
Q(y) = \text{constant} = \frac{1}{y_{\textsc{max}}-y_{\textsc{min}}},
\end{equation}
which is uniform in log-space, meaning that every logarithmic scale represented by a decade is equally likely.

One can now use the probability distribution (\ref{pqy}) which, by virtue Eq. (\ref{1n}), reads:
\begin{equation}
P(x) = \frac{1}{x (y_{\textsc{max}} - y_{\textsc{min}})} = \frac{1}{x \ln(x_{\textsc{max}}/x_{\textsc{min}})}, \quad x \in [x_{\textsc{min}}, x_{\textsc{max}}],
\label{10}\end{equation}
which is also known as the Jeffreys prior in Bayesian statistics. \bltt{We emphasize that Eq.~(\ref{10}) should be interpreted as a scale-invariant reference measure arising from the assumption of scale invariance, and deviations from it should be understood as statistical departures from this reference distribution rather than  thermodynamic or dynamical costs associated with any physical process or as evidence of a dynamically selected equilibrium state. In this work, all entropic differences are therefore interpreted exclusively within this statistical-reference framework.}
Under the assumption of scale invariance, Eq. (\ref{10}) is the only probability distribution that maximizes \sie. The complete formal proof is provided in Appendix \ref{a11}. 
\bltt{As it will be explicit throughout the text, we emphasize that the maximum-entropy characterization of Eq. (10) is conditional on scale invariance and the imposed normalization interval. It does not imply that physical systems necessarily realize this distribution dynamically.}

On a compact log-interval $[y_{\textsc{min}}, y_{\textsc{max}}]$, 
maximizing Shannon information entropy yields a uniform density in the $y$ variable, as 
\begin{align}\label{dely}
S[Q] = - \int_{y_{\textsc{min}}}^{y_{\textsc{max}}} Q(y) \ln Q(y) \, dy 
= \ln (y_{\textsc{max}} - y_{\textsc{min}}),
\end{align}
which is invariant under the shifting by $\ln \lambda$.  

Now, any nonzero real number can be written as $x = m \times 10^k$, for $m \in [1,10)$.
The probability density for the mantissa $m$ follows from integrating $P(x)$ over all decades, $
P(m) \propto \int P(m\times 10^k)\, 10^k\, dk.$ 
Since $P(x) \propto 1/x$ as given by Eq. (\ref{10}), this expression becomes scale independent, and normalization yields  
$P(m) = \frac{1}{m \ln 10}.$
\bltt{Projecting the scale-invariant reference measure onto leading digits yields the standard Benford distribution, which serves as a null hypothesis for scale-free ensembles rather than a dynamical prediction of QCD.}
The probability that the leading digit $d$ attains any value in the set $\{1, 2, \ldots, 9\}$ reads 
\begin{equation}
P(d) = \int_d^{d+1} \frac{1}{m \ln 10}\, dm
= \log_{10}\!\left(1 + \frac{1}{d} \right),\label{nb}
\end{equation}
encoding the Newcomb--Benford's law when projected onto the leading significant digit. Among all possible distributions $P(x)$ on a positive, scale-invariant domain, expression (\ref{nb}) -- and only it -- maximizes \sie.

Consequently, under the assumption of scale invariance, the Shannon 
information entropy is maximized by the log-uniform Newcomb--Benford 
distribution, which represents the most unbiased probability density. 
Scale-invariant distributions describe physical systems with no preferred length or energy scale. In quantum field theory (QFT), scale invariance under 
renormalization group (RG) flow implies that observables exhibit the same 
behavior across energy scales, with no distinguished scale 
\cite{Wilson:1971bg,Polchinski:1983gv}. Although classical QCD with massless 
quarks is scale invariant, quantum corrections break this symmetry through the 
running of the strong coupling, which depends on the renormalization scale 
$\mu$. This introduces the characteristic dimensionful parameter of the theory, 
the QCD scale $\Lambda_{\text{QCD}} \sim 200\text{--}300~\mathrm{MeV}$. For 
$\mu \gg \Lambda_{\text{QCD}}$, the theory lies in its perturbative regime, 
where the running coupling generates asymptotic freedom and quarks and gluons 
interact weakly. On the other hand, the energy range $\mu \lesssim \Lambda_{\text{QCD}}$ 
defines the nonperturbative domain and the coupling becomes strong, giving rise to confinement at low 
energies \cite{GrossWilczek1973,Politzer1973}, manifested in the internal 
structure of hadrons \cite{Duarte:2022yur}. Further details are provided in 
Appendix~\ref{a12}.

Confinement and chiral symmetry breaking generate discrete hadronic mass spectra centered between 140 MeV and 11.1   GeV. 
If the leading-digit distribution of hadronic masses were related to a scale-free ensemble without a preferred energy scale, \sie~would be maximized by the  Newcomb--Benford statistics.
In contrast, QCD dynamics introduce strong correlations and a preferred scale, as hadronization occurs near $\Lambda_{\text{QCD}}$ \cite{Torrieri:2010py,Fontoura:2025fsv}. \bltt{This emergent scale is reflected, at the level of the experimental hadronic mass spectrum, in a reduction of Shannon information entropy relative to the scale-invariant reference ensemble, producing an information-theoretic deficit.} Also, as it will be shown in Section \ref{hip}, it causes the leading-digit distribution of hadron masses to deviate from Newcomb--Benford’s probability distribution, thus signaling the dynamical scale in QCD. 
 We will show that the \sie\ of the hadronic mass spectrum reported by the PDG \cite{pdg} deviates from the maximal (scale-invariant) value, reflecting quantum corrections in QCD that break scale invariance in the IR regime. 

\bltt{All entropy differences to be  introduced in this work should be interpreted strictly as information-theoretic measures of deviation from a scale-invariant reference distribution. In particular, the terms entropy deficit and related quantities do not correspond to thermodynamic entropy production or dynamical work associated with symmetry breaking. Instead, they quantify how the experimental hadronic mass spectrum departs from the maximum-entropy distribution compatible with scale invariance under the chosen normalization and coarse-graining procedure. The information-theoretic framework  establishes that scale invariance leads to a logarithmically uniform probability distribution, maximizing Shannon information entropy and yielding the Newcomb--Benford law as its leading-digit projection. It is therefore natural to ask how the physical hadronic spectrum generated by QCD relates to this idealized scale-invariant reference.}

\bltt{As detailed in Appendix~\ref{a12}, the QCD spectrum emerges from the interplay between confinement and the anomalous breaking of classical scale invariance. The trace anomaly introduces the characteristic scale $\Lambda_{\textsc{QCD}}$, which simultaneously generates a mass gap in the IR and organizes the tower of hadronic excitations. A key empirical consequence is the approximate exponential growth of the density of states,
\begin{equation}
\rho(M) \propto e^{M/T_{\textsc{H}}},
\end{equation}
where $T_\textsc{H}$ is the Hagedorn temperature, 
implying that the cumulative number of states behaves as $N(M)\sim e^{M/T_{\textsc{H}}}$ \cite{Hagedorn1968,Hagedorn1965,Cabibbo:1975ig,GrossWilczek1973,Politzer1973}. 
This structure has a direct interpretation in terms of information theory. An exponential distribution in the mass variable $M$ corresponds to an approximately uniform distribution in $\ln M$, which is precisely the configuration that maximizes Shannon entropy under scale invariance, as encoded in Eq. (\ref{dely}). In this sense, the use of logarithmic variables is not merely a convenient choice, but reflects the intrinsic scaling properties of the QCD spectrum. 
However, QCD is not exactly scale invariant. The presence of the mass gap, the confinement scale, and the transition between light- and heavy-flavor sectors introduce nontrivial structure in the spectrum, breaking the ideal logarithmic uniformity. These effects distort the probability distribution away from the scale-invariant reference measure $P(x)\propto 1/x$, and consequently reduce the associated Shannon information entropy.}

\bltt{Therefore, the entropy measures introduced in this work quantify how the experimental hadronic mass distribution departs from the maximum-entropy configuration dictated by scale invariance. In particular, deviations from the Newcomb--Benford distribution should be interpreted as signatures of the dynamical scales inherent to QCD, rather than as violations of a purely statistical law. This establishes a direct conceptual link between the emergence of the QCD spectrum and the information-theoretic framework employed throughout this work.}

\section{\sie~for Hadrons}
\label{hip}
Our analysis is organized into three cases, in which the \sie~of the 
leading-digit distribution of the hadronic mass spectrum is computed and 
compared with the \sie~associated with the Newcomb--Benford law. As will be 
shown, the latter is statistically suppressed owing to the presence of a 
preferred energy scale in QCD. Subsec.~\ref{mfs} examines the meson states 
listed in the PDG, Subsec.~\ref{bh} treats the baryon sector, and 
Subsec.~\ref{mfsb} considers the combined meson--baryon spectrum \cite{pdg}. 
For definiteness, let the leading digit of a hadron mass be $d$, with 
$\mathcal{O}_d$ denoting the number of observed occurrences and 
$P_{\textsc{obs}}(d)=\mathcal{O}_d/N$ the corresponding empirical probability, where $N$ denotes the total number in the hadronic sample. The expected 
count under the Newcomb--Benford law is 
$\mathcal{E}_d = N\,P_{\textsc{Benf}}(d)$, where $P_{\textsc{Benf}}(d)$ is the 
associated theoretical probability distribution.
The mass spectra of mesons and baryons are presented combined  (Fig. \ref{fig:10}) and  separately (Fig. \ref{fig:20}). 
\begin{figure}[H]
	\centering
	\includegraphics[width=11cm]{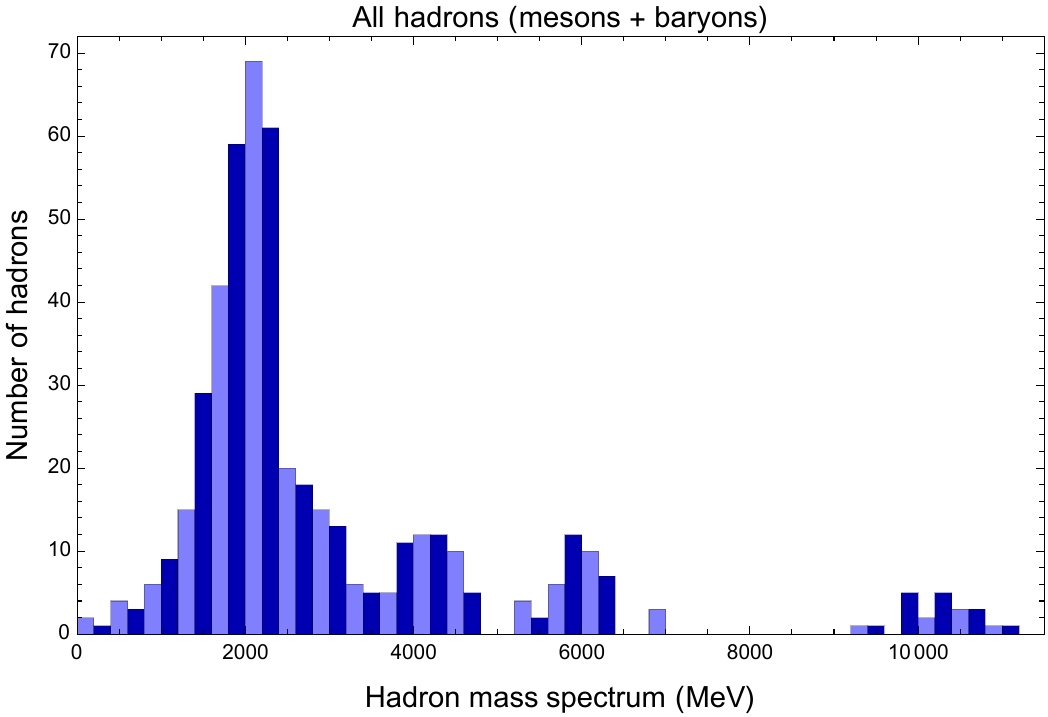}
	\caption{Distribution of hadronic mass spectrum in PDG \cite{pdg}.}
	\label{fig:10}
\end{figure}

\begin{figure}[H]
	\centering
	\!\!\!\!\!\!\!\!\!\!\!\!\!\!\!\includegraphics[width=17cm]{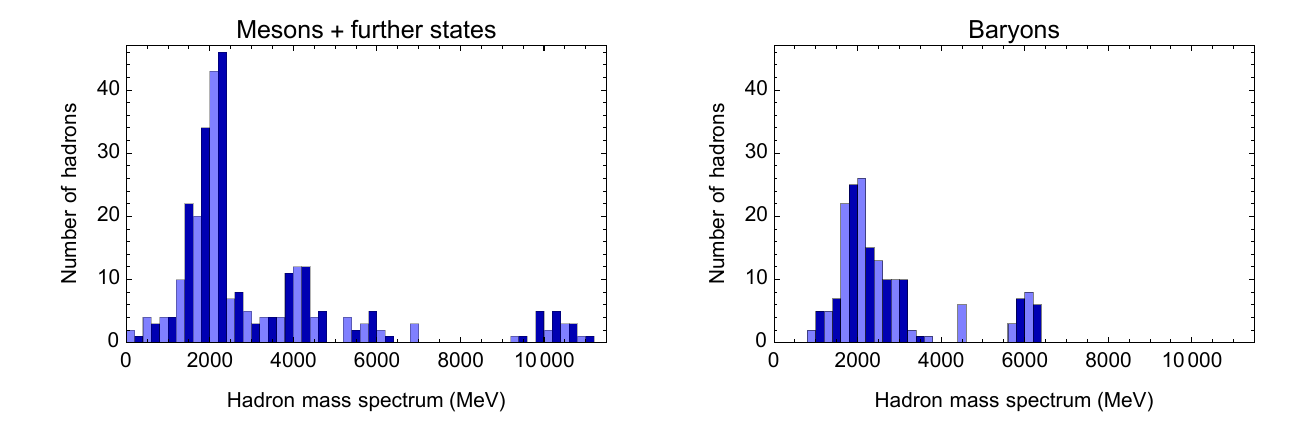}
	\caption{Distribution of meson (with further mesonic states) [left panel] and baryon [right panel] mass spectrum in PDG \cite{pdg}.}
	\label{fig:20}
\end{figure}

\subsection{\sie~for mesons and further mesonic states in PDG}
\label{mfs}
We analyze the distribution of the meson mass spectrum, as listed in the PDG \cite{pdg}, whose leading digit counts are displayed in Table~\ref{t259}. 
\bltt{We note that the hadronic mass inputs from the PDG include both narrow states and broad resonances, many of which are characterized by finite decay widths rather than sharply defined masses. In the present analysis, each resonance is represented by its quoted central mass value \cite{pdg}. This introduces a potential source of systematic uncertainty, since the true physical state corresponds to a mass distribution whose leading digit may fluctuate depending on the precise definition of the representative mass as, for instance, the pole mass, the Breit--Wigner peak, or an averaged value \cite{pdg}. A more refined treatment could incorporate resonance widths through a smearing procedure or Monte Carlo sampling of the mass distribution to assess the stability of the leading-digit statistics under experimental resolution effects.
In addition, it is important to note that the PDG hadronic spectrum reports a carefully curated experimental compilation of QCD states, rather than a fully exhaustive or uniformly sampled realization of the underlying theory. It reflects the current state of experimental accessibility, with particularly rich coverage in the low-mass and light-quark sectors, while progressively more sparse information is available for heavy-flavor and highly excited resonances. Rather than reflecting a shortcoming, this structure encodes the current frontier of hadron spectroscopy and naturally shapes the empirical leading-digit distribution and associated information-theoretic measures, including information entropy-based measures. The results presented here should therefore be interpreted as properties of the observed hadronic spectrum rather than of the full underlying QCD state space.}
\begin{table}[H]
\centering
\begin{tabular}{|c||S[table-format=3.0]|S[table-format=1.3]!{\vrule width 1pt}|S[table-format=3.1]|S[table-format=1.3]|}
\hline
{Digit ($d$)} & {{Observed ($\mathcal{O}_d$})} & {{ $P_{\textsc{obs}}(d)$}} 
& {{Benford ($\mathcal{E}_d$)}} & {{$P_{\textsc{Benf}}(d)$}} \\
\hline\hline
1 & 112 & 0.333 & 101.2 & 0.301 \\
2 & 123 & 0.366 & 59.2  & 0.176 \\
3 & 30  & 0.089 & 42.0  & 0.125 \\
4 & 35  & 0.104 & 32.6  & 0.097 \\
5 & 16  & 0.048 & 26.6  & 0.079 \\
6 & 6   & 0.018 & 22.5  & 0.067 \\
7 & 3   & 0.009 & 19.5  & 0.058 \\
8 & 1   & 0.003 & 17.1  & 0.051 \\
9 & 10  & 0.030 & 15.4  & 0.046 \\
\hline
\end{tabular}
\caption{Comparison of the leading-digit distribution of the  meson mass spectrum (with further mesonic states) in PDG to Newcomb--Benford's probability distribution ($N = 336$).}\label{t259}
\end{table}
\begin{figure}[H]
	\centering
	\includegraphics[width=10cm]{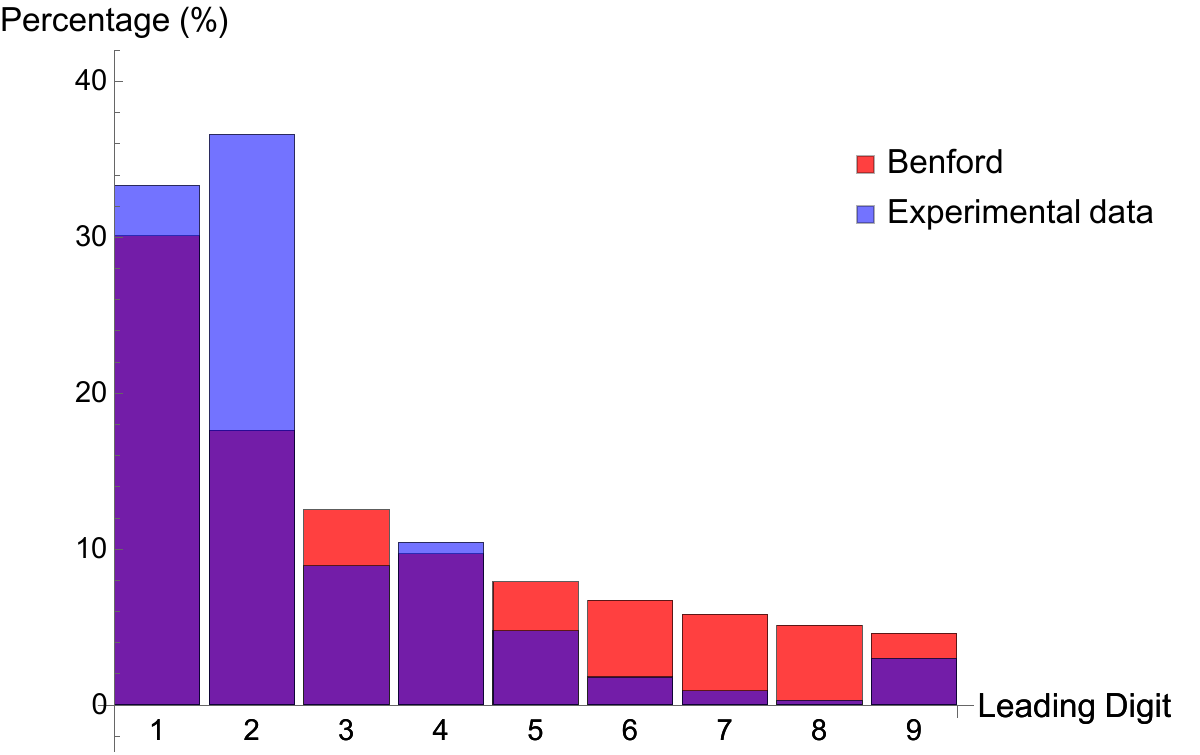}
	\caption{Distribution of the meson mass spectrum with further mesonic states in PDG \cite{pdg},  according to the leading digit.}
	\label{fig:2mf}
\end{figure}

Shannon information entropy of the observed meson mass spectrum can be calculated by taking the dataset at the third column of Table \ref{t259}, reading
\begin{align}
S_{\textsc{PDG}} &= -\sum_{d=1}^9 P_{\textsc{obs}}(d)\ln P_{\textsc{obs}}(d) = 1.569~\text{nat}. 
\end{align} The Shannon information entropy for the Newcomb--Benford distribution can be computed when addressing the dataset at the fifth column of Table \ref{t259}, yielding  
\begin{align}
S_{\textsc{Benf}} &= -\sum_{d=1}^9 P_{\textsc{Benf}}(d)\ln P_{\textsc{Benf}}(d) = 1.996~\text{nat}.
\end{align}The Shannon information entropy deficit reads 
\begin{equation}\label{ed1}
\Delta S = S_{\textsc{Benf}} - S_{\textsc{obs}} = 0.427~\text{nat},
\end{equation}
\bltt{corresponding to a 21.37\% deviation from the scale-invariant reference distribution. This result is consistent with a characteristic hadronic scale in QCD, which induces departures from scale invariance, and is compatible with QCD-driven scale-setting effects within statistical uncertainties.}

To quantify the deviation from Newcomb--Benford's law, a $\chi^2$ test can be implemented from the meson mass spectrum dataset in Table \ref{t259}, as 
\begin{equation}
\chi^2_{\textsc{obs}} = \sum_{d=1}^{9} \frac{(\mathcal{O}_d - \mathcal{E}_d)^2}{\mathcal{E}_d} = \sum_{d=1}^{9} \frac{(N P_{\textsc{obs}} - N P_{\textsc{Benf}})^2}{N P_{\textsc{Benf}}}=179.859,
\end{equation}
whose $p$-value at the 5\% significance level reads 
\begin{equation}\label{p}
p \;=\;
\frac{\Gamma\!\left(4,\, {\chi_{\mathrm{obs}}^{2}}/{2}\right)}
     {\Gamma\!\left(4\right)}\,,
\end{equation} where $\Gamma(s)$ is the gamma function, and $\Gamma(s,x)=\int_{x}^{\infty}t^{s-1}\,e^{-t}\,dt$ is the upper incomplete gamma function. Eq. (\ref{p}) corresponds to the probability of observing a 
$\chi^2$ distribution value as large or larger than the observed one under the null hypothesis -- that the leading-digit distribution follows Newcomb--Benford's law -- reads 
$p = P(\chi^2\geq 179.859)=1.075\times 10^{-34}$.
\bltt{The small $p$-value indicates statistically significant deviation from the Benford reference distribution within the present dataset.}

It supports QCD-induced energy scale effects and clustering in the meson mass spectrum. 
 The emergence of $\Lambda_{\text{QCD}}$  breaks scale invariance and lowers  \sie~relative to a scale-free ensemble.
Therefore, the leading-digit distribution of the meson mass spectrum and its \sie~reflect the existence of an energy scale in QCD. \bltt{The entropy deficit provides a quantitative measure of deviation from the scale-invariant reference distribution in the hadronic spectrum, rather than a thermodynamic or energy cost associated with symmetry breaking.}

\subsection{\sie~for baryons in PDG}
\label{bh}

The leading-digit statistics of the baryon mass spectrum can be analogously analyzed, also showing scale invariance breaking in the baryonic sector. As also holds for mesons, if scale invariance held, the leading-digit distribution of the baryonic masses would follow Newcomb--Benford's law. Deviations from it, therefore, quantify the degree to which QCD dynamics and confinement effects introduce preferred mass scales into the baryon spectrum. The observed and expected probabilities for each leading digit are summarized in Table~\ref{3376}. 
\begin{table}[H]
\centering
\begin{tabular}{|c||S[table-format=3.0]|S[table-format=1.3]!{\vrule width 1pt}|S[table-format=3.1]|S[table-format=1.3]|}
\hline
{Digit ($d$)} & {{Observed ($\mathcal{O}_d$)}} & {{ $P_{\textsc{obs}}(d)$}} 
& {{Benford ($\mathcal{E}_d$)}} & {{$P_{\textsc{Benf}}(d)$}} \\
\hline\hline
1 & 64 & 0.339 & 56.8 & 0.301 \\
2 & 74 & 0.392 & 33.3 & 0.176 \\
3 & 15 & 0.079 & 23.6 & 0.125 \\
4 & 6  & 0.032 & 18.3 & 0.097 \\
5 & 10 & 0.053 & 14.9 & 0.079 \\
6 & 18 & 0.095 & 12.6 & 0.067 \\
7 & 0  & 0.000 & 11.0 & 0.058 \\
8 & 0  & 0.000 & 9.7  & 0.051 \\
9 & 2  & 0.011 & 8.7  & 0.046 \\\hline
\end{tabular}
\caption{Comparison of the leading-digit distribution of baryon mass spectrum in PDG with the Newcomb--Benford's probability distribution ($N = 189$).}\label{3376}
\end{table}
Fig. \ref{fig:2b} illustrates the observed baryon mass spectrum. 
\begin{figure}[H]
	\centering
	\includegraphics[width=10cm]{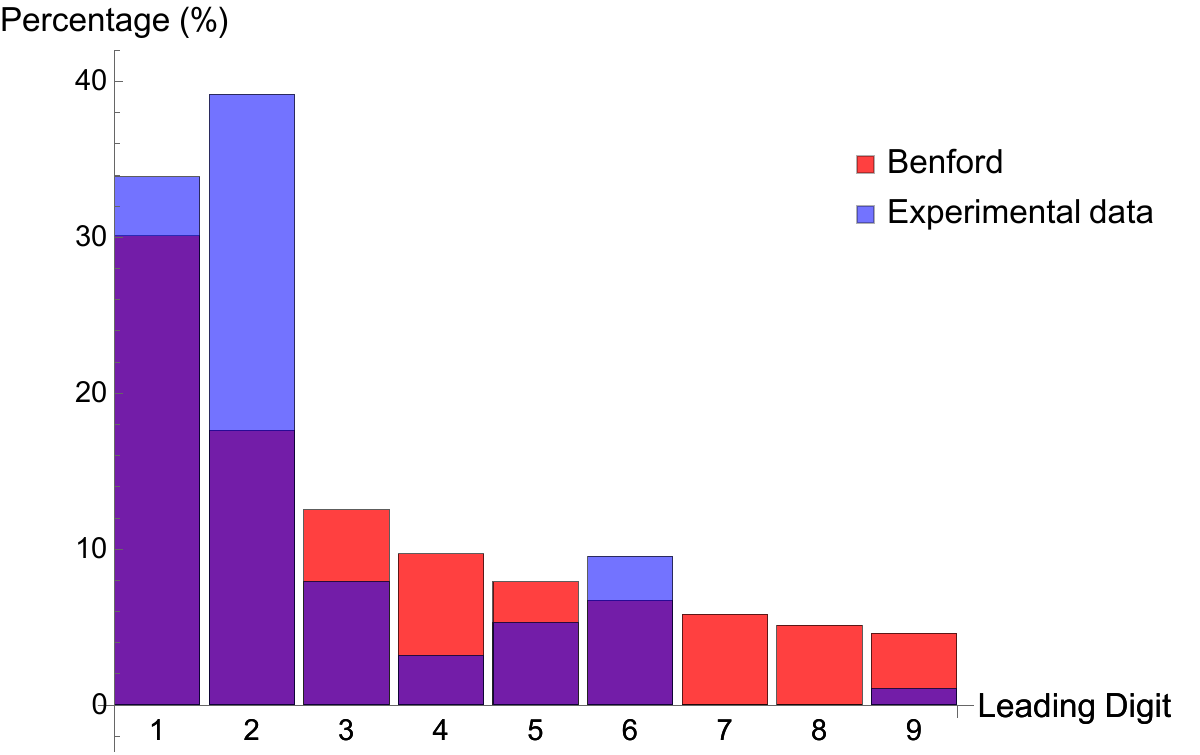}
	\caption{Distribution of the baryon mass spectrum in PDG \cite{pdg},  according to the leading digit.}
	\label{fig:2b}
\end{figure}
Shannon information entropy of the observed meson mass spectrum can be calculated by taking the dataset in the third column of Table \ref{3376}, reading
\begin{align}
S_{\textsc{PDG}} &= -\sum_{d=1}^9 P_{\textsc{obs}}(d)\ln P_{\textsc{obs}}(d) = 1.472~\text{nat}. 
\end{align} Shannon information entropy for the Newcomb--Benford distribution can also be computed when addressing the dataset at the fifth column of Table \ref{3376}, yielding  
\begin{align}
S_{\textsc{Benf}} &= -\sum_{d=1}^9 P_{\textsc{Benf}}(d)\ln P_{\textsc{Benf}}(d) = 1.996~\text{nat}.
\end{align}The Shannon information entropy deficit reads 
\begin{equation}\label{5240}
\Delta S = S_{\textsc{Benf}} - S_{\textsc{obs}} = 0.524~\text{nat},
\end{equation}
\bltt{corresponding to a reduction relative to the scale-invariant reference distribution, consistent with the presence of QCD-induced structure in the experimental mass spectrum.}
From a purely statistical viewpoint, Newcomb--Benford’s law corresponds to the maximum-\sie~distribution of leading digits under scale invariance, where all logarithmic decades are equally probable, and no physical scale dominates. 
However, the leading-digit distribution of the baryon mass spectrum in PDG reveals a \sie~deficit (\ref{5240}), meaning that the digit distribution carries \bltt{measurable statistical deviation from a scale-invariant reference, associated with the characteristic mass scale $\Lambda_{\textsc{QCD}}$.} 
The $\chi^2$ test yields $\chi^2_{\textsc{obs}}=124.320$, with $p$-value given by  $
p = P(\chi^2 \ge 124.320) \lesssim 4.241 \times 10^{-23}$. \bltt{The small $p$-value indicates statistical incompatibility with the Benford null hypothesis for this dataset.}

\subsection{\sie~for mesons (with further mesonic states) and baryons in PDG}
\label{mfsb}
The leading-digit statistics of meson and baryon mass spectra will be investigated. Table~\ref{525} reports the observed and expected frequencies, together with the  probabilities from Newcomb--Benford’s law for $N = 525$ entries. 
\begin{table}[H]
\centering
\begin{tabular}{|c||S[table-format=3.0]|S[table-format=1.3]!{\vrule width 1pt}|S[table-format=3.1]|S[table-format=1.3]|}
\hline
{Digit ($d$)} & {{Observed ($\mathcal{O}_d$)}} & {{ $P_{\textsc{obs}}(d)$}} 
& {{Benford ($\mathcal{E}_d$)}} & {{$P_{\textsc{Benf}}(d)$}} \\
\hline\hline
1 & 176 & 0.335 & 157.8 & 0.301 \\
2 & 197 & 0.375 & 92.4  & 0.176 \\
3 & 45  & 0.086 & 65.6  & 0.125 \\
4 & 41  & 0.078 & 50.9  & 0.097 \\
5 & 26  & 0.050 & 41.6  & 0.079 \\
6 & 24  & 0.046 & 35.1  & 0.067 \\
7 & 3   & 0.006 & 30.4  & 0.058 \\
8 & 1   & 0.002 & 26.8  & 0.051 \\
9 & 12  & 0.023 & 24.0  & 0.046 \\
\hline
\end{tabular}
\caption{Comparison of the leading-digit distribution of meson (with further mesonic states) and baryon masses in PDG to the Newcomb--Benford's probability distribution ($N = 525$).}\label{525}
\end{table}
Fig. \ref{fig:3c} depicts the observed hadronic mass spectrum.
\begin{figure}[H]
	\centering
	\includegraphics[width=10cm]{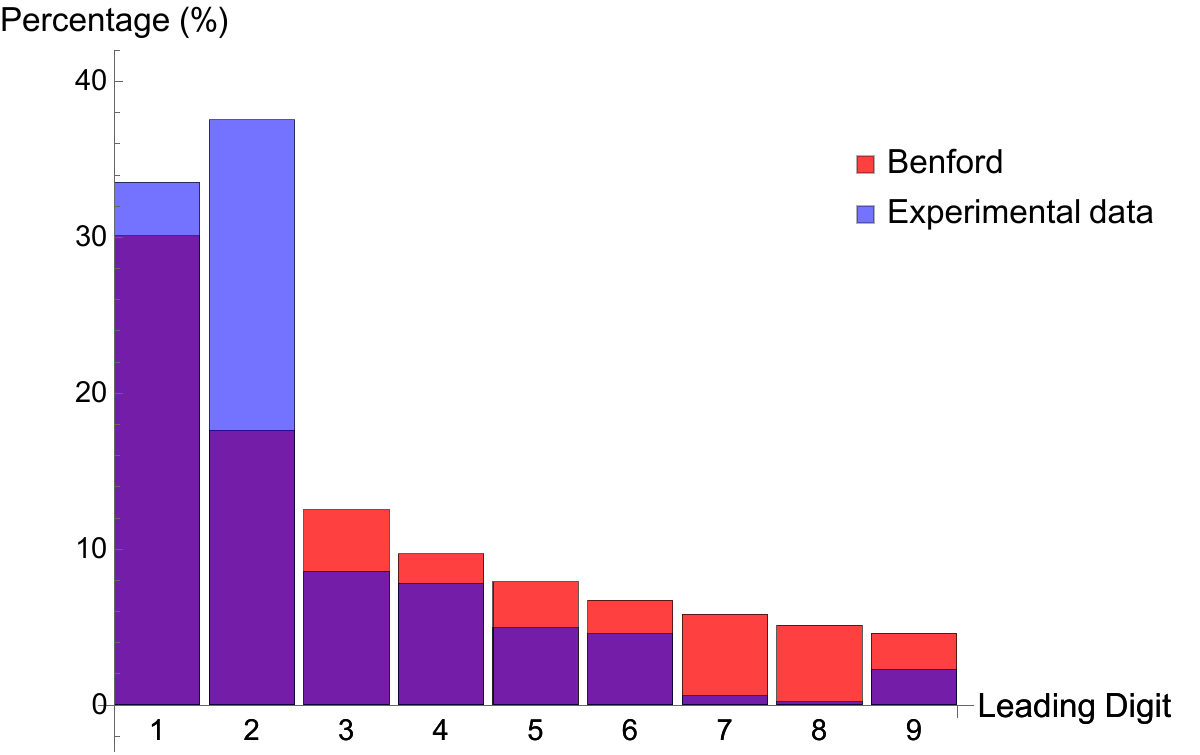}
	\caption{Distribution of the meson mass spectrum with further mesonic states  and baryons in PDG \cite{pdg}, according to the leading digit.}
	\label{fig:3c}
\end{figure}
Digits 1 and 2 are overrepresented, while higher numbers are underrepresented relative to the ideal logarithmic Newcomb--Benford distribution, although clustering is again observed in the leading digit 9. 
Shannon information entropy of the observed baryon  mass spectrum can be calculated by the third column of Table \ref{525}, yielding
\begin{align}
S_{\textsc{PDG}} &= -\sum_{d=1}^9 P_{\textsc{obs}}(d)\ln P_{\textsc{obs}}(d) = 1.566~\text{nat}. 
\end{align} The Shannon information entropy for the Newcomb--Benford distribution can be computed when addressing the dataset at the fifth column of Table \ref{525}, yielding  
\begin{align}
S_{\textsc{Benf}} &= -\sum_{d=1}^9 P_{\textsc{Benf}}(d)\ln P_{\textsc{Benf}}(d) = 1.996~\text{nat},
\end{align}implying the Shannon information entropy deficit:
\begin{equation}
\Delta S = S_{\textsc{Benf}} - S_{\textsc{obs}} = 0.430~\text{nat},\label{ed11}
\end{equation}
representing a $21.55\%$ reduction. 
Analogously to what has been reported for baryons and mesons separately in the previous subsections, the combined mesonic and baryonic mass spectra show that $\chi_{\textsc{obs}}^2 = 329.525$. 
The $p$-value $
p = P(\chi^2 \ge 329.525)\lesssim 2.110 \times 10^{-66}$ is 
infinitesimal. \bltt{It suggests rejection of the Benford hypothesis at conventional significance levels.} 
 The deviation is primarily due to the overpopulation of digits 1 and 2.
The \sie~deficit (\ref{ed11}) quantifies the information gained from QCD dynamical emergence of the scale $\Lambda_{\text{QCD}}$. Thus, the combined meson and baryon leading-digit statistics \bltt{provides an information-theoretic characterization of deviations from scale invariance associated with the presence of the QCD scale $\Lambda_{\text{QCD}}$.} The hadronic mass spectrum clusters around certain mass ranges,  suppressing the logarithmic uniformity in the hadronic mass leading digit.


\section{Additional information-theoretic measures: KL divergence, Jensen--Shannon divergence, conditional and mutual entropies}
\label{sec:info}

The Shannon information entropy provides a first quantitative indication of
scale-invariance breaking in QCD through deviations from the  
Newcomb--Benford distribution.  Besides, other information-theoretic measures
offer complementary and, in some cases, sharper measures of the  
degree to which the hadronic mass spectrum departs from logarithmic 
scale invariance.  In this section, we evaluate the 
Kullback--Leibler divergence, the Jensen--Shannon divergence, the conditional
entropy of the leading digit given the hadronic sector, and the mutual
information between these variables \cite{Witten:2018zva}.  These quantities probe the 
information-theoretic imprint of $\Lambda_{\textsc{QCD}}$ in the hadron
spectrum beyond the Shannon entropy deficit.  
All computations use the experimental digit probabilities listed in 
Tables~\ref{t259}, \ref{3376}, and \ref{525}. \bltt{We stress that the following constructions are classical information-theoretic analogues of entanglement entropy, as no underlying quantum bipartition is assumed.}

\subsection{Kullback--Leibler divergence}

Given an observed leading-digit distribution $P_{\textsc{obs}}(d)$
and the Benford distribution $P_{\textsc{Benf}}(d)$, the 
Kullback--Leibler (KL) divergence is defined by
\begin{equation}
D_{\mathrm{KL}}(P_{\textsc{obs}}\|P_{\textsc{Benf}})
=\sum_{d=1}^9 P_{\textsc{obs}}(d)\,
\ln\!\frac{P_{\textsc{obs}}(d)}{P_{\textsc{Benf}}(d)}\,.
\end{equation}
$D_{\mathrm{KL}}=0$ only if the observed distribution exactly matches the
scale-invariant Benford form.  \bltt{The Kullback--Leibler divergence quantifies the information-theoretic distance between the experimental distribution and the scale-invariant reference distribution.}

{}{Evaluating the Kullback--Leibler divergence for the three datasets yields
\begin{align}
D_{\mathrm{KL}}^{\textsc{mesons}}
&= 0.134\ \text{nat},\\
D_{\mathrm{KL}}^{\textsc{baryons}}
&= 0.218\ \text{nat},\\
D_{\mathrm{KL}}^{\textsc{combined}}
&= 0.139\ \text{nat}.
\end{align}
The KL divergence is largest in the baryonic sector, indicating that baryon masses exhibit the strongest departure from the logarithmic,
scale-invariant behaviour. \bltt{These divergences quantify distributional differences but do not by themselves establish a dynamical origin.} This is consistent with the larger dynamical
mass generated in three-quark bound states.}
\bltt{It should be noted that the numerical value of $D_{\mathrm{KL}}$ depends on the finite experimental sampling of the hadronic spectrum in PDG \cite{pdg} and on the induced coarse-graining associated with mapping masses into leading-digit classes, and therefore reflects a discretized comparison between distributions.}

\subsection{Jensen--Shannon divergence}
{}{
A symmetric measure of the similarity between two probability distributions is the Jensen--Shannon divergence (JSD), defined as
\begin{equation}
\mathrm{JSD}(P_{\textsc{obs}},P_{\textsc{Benf}})
=\frac12 D_{\mathrm{KL}}(P_{\textsc{obs}}\|M)
+ \frac12 D_{\mathrm{KL}}(P_{\textsc{Benf}}\|M),
\qquad
M = \frac12 (P_{\textsc{obs}} + P_{\textsc{Benf}}).
\end{equation}
The mixture distribution $M(d)$ for mesons is given in Table~\ref{tab:M_mesons}.
\begin{table}[H]
\centering
\begin{tabular}{|c||S[table-format=1.3]|S[table-format=1.3]!{\vrule width 1pt}|S[table-format=1.3]|}
\hline
{Digit ($d$)} & {{$P_{\textsc{obs}}(d)$}} & {{$P_{\textsc{Benf}}(d)$}} 
& {{$M(d) = \frac{P_{\textsc{obs}}+P_{\textsc{Benf}}}{2}$}} \\
\hline\hline
1 & 0.333 & 0.301 & 0.317 \\
2 & 0.366 & 0.176 & 0.271 \\
3 & 0.089 & 0.125 & 0.107 \\
4 & 0.104 & 0.097 & 0.101 \\
5 & 0.048 & 0.079 & 0.064 \\
6 & 0.018 & 0.067 & 0.043 \\
7 & 0.009 & 0.058 & 0.034 \\
8 & 0.003 & 0.051 & 0.027 \\
9 & 0.030 & 0.046 & 0.038 \\
\hline
\end{tabular}
\caption{The mixture distribution $M(d)$ used in the Jensen--Shannon divergence calculation for the meson mass spectrum in PDG \cite{pdg}.}
\label{tab:M_mesons}
\end{table}
Numerical evaluations of the JSD yield
\begin{align}
\mathrm{JSD}_{\textsc{mesons}} &= 0.0276~\text{nat},\\
\mathrm{JSD}_{\textsc{baryons}} &= 0.0473~\text{nat},\\
\mathrm{JSD}_{\textsc{combined}} &= 0.0285~\text{nat}.
\end{align}
The hierarchy mirrors that of the KL divergence, confirming that baryons are furthest from the scale-invariant logarithmic distribution.}
\bltt{The reported JSD likewise depends on the adopted coarse-graining procedure and should be interpreted as a stable yet coarse-grained measure of similarity between the underlying distributions.}

\subsection{Conditional entropy and mutual information}

{}{Let $S \in \{M,B\}$ denote the hadronic sector. The conditional entropy of the leading digit given the sector is
\begin{equation}
H(D|S) = p_{M} H(D|M) + p_{B} H(D|B),
\end{equation}
where $p_M = 336/525 = 0.640$ and $p_B = 189/525 = 0.360$ are the fractions of mesons and baryons in the combined dataset, respectively. Using the previously computed entropies for each sector,
\begin{align}
H(D|M) &= -\sum_{d=1}^9 P_{\textsc{obs}}^{\textsc{mesons}}(d)\ln P_{\textsc{obs}}^{\textsc{mesons}}(d) = 1.569~\text{nat},\\
H(D|B) &= -\sum_{d=1}^9 P_{\textsc{obs}}^{\textsc{baryons}}(d)\ln P_{\textsc{obs}}^{\textsc{baryons}}(d) = 1.472~\text{nat},\\
H(D|S) &= 0.640 \times 1.569 + 0.360 \times 1.472 = 1.533~\text{nat}.
\end{align}
The mutual information between $D$ and $S$ quantifies the reduction in uncertainty about the leading digit when the sector is known:
\begin{equation}
I(D;S) = H(D) - H(D|S) = 1.566 - 1.533 = 0.033~\text{nat}.
\end{equation}
Although modest, this positive mutual information indicates that knowledge of whether a hadron is a meson or baryon slightly increases the predictability of its leading digit, consistent with the sector-dependent imprint of $\Lambda_{\textsc{QCD}}$.}

{}{The collection of information-theoretic quantities 
$\big\{D_{\mathrm{KL}},\mathrm{JSD},H(D|S),I(D;S)\big\}$ \bltt{all indicate a consistent statistical deviation from the scale-invariant reference distribution, reflecting the presence of QCD-induced structure in the hadronic mass spectrum.}  
The baryonic sector, in particular, lies farther from the 
scale-invariant Benford fixed point than the mesonic sector, 
as encoded by the larger divergences and the lower conditional entropy.  
The nonvanishing mutual information between the leading digit and the 
hadron sector confirms that QCD dynamics produce distinguishable 
information-theoretic patterns in different classes of hadrons.  
While none of these measures determines the numerical value of 
$\Lambda_{\textsc{QCD}}$, they quantify the extent to which its 
emergence constrains and structures the statistical organization of the
hadronic mass spectrum.}
\bltt{The mutual information depends on the discretized representation of the mass spectrum and should be interpreted as a measure of statistical dependence induced by coarse-graining, rather than a fundamental dynamical correlation. All information-theoretic quantities introduced in this work, including mutual information, conditional entropy, entropy deficits, and divergence measures, are computed from classical probability distributions derived from hadronic mass spectra. While some notation borrows terminology from quantum information theory, this is purely a formal analogy reflecting structural similarities between bipartite probability decompositions and quantum partial traces. No quantum state, Hilbert space structure, or physical entanglement is assumed, and the density matrices employed are outer-product representations of probability vectors used as auxiliary constructs. Accordingly, all entropies are standard Shannon entropies of classical distributions. These quantities should be understood as measures of deviation from a scale-invariant reference ensemble, rather than as thermodynamic energy costs or dynamical work associated with breaking scale invariance. Instead, they quantify the extent to which the experimental hadronic mass spectrum departs from a maximum-entropy, scale-invariant distribution.}

\section{Information-theoretic bipartitions of the hadronic spectrum}
\label{sec:entanglement}

{}{The Shannon, KL, and JS information entropies quantify the 
departure of the experimental leading-digit distribution from the scale-invariant 
Newcomb--Benford law.  These quantities probe global deviations from scale 
invariance, but one can also construct information-theoretic measures that 
characterize correlations {between} different degrees of freedom in the 
hadronic spectrum.  While the hadronic mass spectra in PDG do not provide direct access to 
quantum-mechanical density matrices from which genuine von~Neumann 
entanglement may be computed, it is nevertheless meaningful to construct 
entanglement-entropy-like quantities in the classical-information sense.  
Such constructions are widely used in statistical physics and nonlinear dynamics
to quantify bipartite correlations in systems lacking an underlying Hilbert-space description \cite{Casini:2011kv,daRocha:2020gee}.  }
\bltt{The use of entanglement-entropy language is purely formal. All quantities are defined from classical probability distributions of hadronic mass spectra and do not imply quantum entanglement, density matrices from quantum states, or any Hilbert-space structure. The terminology is used strictly as an analogy to emphasize structural similarities with bipartite information measures in quantum information theory.}

\bltt{\subsection{Classical bipartite entropy and sector-digit correlations}}
{}{Let $D$ denote the leading digit of a hadron mass and $S$ the hadronic sector 
(meson or baryon).  The experimental joint distribution $P(d,s)$ over the $9\times 2$
possibilities defines a classical \bltt{joint probability distribution}
\begin{equation}
\rho_{DS}(d,s) = P(d,s),
\end{equation}
from which \bltt{one obtains marginal probability distributions} $\rho_D$ and $\rho_S$.  
The Shannon entropies $H(D)$, $H(S)$, and $H(D,S)$ then define the classical 
mutual information,
\begin{equation}
I(D;S) = H(D) + H(S) - H(D,S).
\end{equation}
The joint probability distribution of leading digit $D$ and hadron sector $S$ (meson $M$, baryon $B$) is obtained from the combined PDG dataset in Table~\ref{525}:
\begin{equation}
\rho_{DS}(d,s) = P(D=d, S=s) = \frac{\mathcal{O}_{d,s}}{N}, \qquad N=525\,.
\end{equation}
\begin{table}[H]
\centering
\begin{tabular}{|c||c|c|}
\hline
Digit $d$ & $\rho_{DS}(d,M)$ & $\rho_{DS}(d,B)$ \\
\hline
1 & 0.2133 & 0.1219 \\
2 & 0.2343 & 0.1410 \\
3 & 0.0571 & 0.0286 \\
4 & 0.0667 & 0.0114 \\
5 & 0.0305 & 0.0190 \\
6 & 0.0114 & 0.0343 \\
7 & 0.0057 & 0.0000 \\
8 & 0.0019 & 0.0000 \\
9 & 0.0190 & 0.0038 \\
\hline
\end{tabular}
\caption{Joint probability distribution $\rho_{DS}(d,s)$ for leading digit $d$ and sector $s$.}
\label{tab:joint_DS}
\end{table}
The marginals are then given by 
\begin{align}
\rho_D(d) &= \sum_{s \in \{M,B\}} \rho_{DS}(d,s), \\
\rho_S(s) &= \sum_{d=1}^9 \rho_{DS}(d,s).
\end{align}
respectively presented in Tables \ref{tab:marginal_D} and \ref{tab:marginal_S}.
\begin{table}[H]
\centering
\begin{tabular}{|c||S[table-format=1.0]|S[table-format=1.4]|}
\hline
{Digit ($d$)} & {{Observed ($\mathcal{O}_d$)}} & {{$\rho_D(d)$}} \\
\hline\hline
1 & 176 & 0.3352 \\
2 & 197 & 0.3753 \\
3 &  45 & 0.0857 \\
4 &  41 & 0.0781 \\
5 &  26 & 0.0495 \\
6 &  24 & 0.0457 \\
7 &   3 & 0.0057 \\
8 &   1 & 0.0019 \\
9 &  12 & 0.0228 \\
\hline
\end{tabular}
\caption{Leading-digit marginal probabilities $\rho_D(d)$.}
\label{tab:marginal_D}
\end{table}
\begin{table}[H]
\centering
\begin{tabular}{|c||c|}
\hline
Sector $s$ & $\rho_S(s)$ \\
\hline
Meson $M$ & 0.64 \\
Baryon $B$ & 0.36 \\
\hline
\end{tabular}
\caption{Sector marginal probabilities $\rho_S(s)$.}
\label{tab:marginal_S}
\end{table}
The mutual information entropy for the combined meson and baryon dataset is given by 
\begin{eqnarray}
H(D) &=& - \sum_{d=1}^9 \rho_D(d) \ln \rho_D(d) = 1.566~\text{nat},\\
H(S) &=& - \sum_{s \in \{M,B\}} \rho_S(s) \ln \rho_S(s) = 0.664~\text{nat},\\
H(D,S) &=& - \sum_{d=1}^9 \sum_{s \in \{M,B\}} \rho_{DS}(d,s) \ln \rho_{DS}(d,s) = 2.197~\text{nat},
\end{eqnarray} which implies that 
\begin{eqnarray}
I(D;S) &=& H(D) + H(S) - H(D,S) =0.033~\text{nat}.\label{ids}
\end{eqnarray}
These values indicate that while the leading digit carries most of the information globally ($H(D)= 1.566$), knowledge of the sector slightly reduces uncertainty, yielding a small but nonzero mutual information, $I(D;S)= 0.033$, reflecting the statistical correlation between meson/baryon classification and leading-digit structure in the hadronic mass spectrum.
In classical information theory, this quantity is interpreted as the 
{shared entropy} between the two subsystems and \bltt{serves as a classical analogue of entanglement entropy used in quantum systems}.  Eq. (\ref{ids}) indicates a small but statistically significant correlation between the 
leading-digit structure and the hadronic sector.  
In the present context, this classical entanglement entropy \bltt{measures statistical correlations between the digit structure and the hadronic-sector classification in the experimental dataset}.  
The nonzero value of the mutual information (\ref{ids}) encodes distinct mass-clustering patterns in the mesonic
and baryonic mass spectra.}

\bltt{\subsection{Robustness and interpretation of discretized information measures}}
\label{new1}
\bltt{The information-theoretic quantities, namely the KL divergence, JS divergence, and mutual information, are computed from a discretized representation of the hadronic mass spectrum. As a consequence, they should be interpreted as coarse-grained statistical descriptors of the experimental distribution rather than as fundamental observables of the underlying quantum field theory. 
In particular, these measures depend on the chosen binning scheme used to partition the continuous mass spectrum into discrete categories. While logarithmic binning is physically motivated by the scale-invariant reference measure, alternative reasonable discretizations may lead to quantitatively different values, although the qualitative hierarchy among meson, baryon, and combined datasets is expected to remain stable. 
We therefore emphasize that the results presented here quantify deviations from a scale-invariant reference at the level of statistical structure in the observed spectrum, rather than providing a direct reconstruction of non-perturbative QCD dynamics. Within this interpretation, these measures serve as probes of coarse-grained organization induced by the presence of a characteristic hadronic scale.}
\bigskip

\bltt{\section{Spectral bipartition entropy in a logarithmic bin decomposition}}
\label{sec:UVIR}

{}{An entanglement-entropy-like analysis of scale-invariance breaking in the hadronic mass spectrum of QCD, can be constructed by splitting the hadronic mass spectra into, e.g. $10$ logarithmic bins spanning the hadronic mass range $0.139$ -- $11.1~\mathrm{GeV}$.  
Let $P_k$ denote the normalized population of the bin $k$.  
\textcolor{black}{In this section, all probabilities are normalized using the total number of states in the full dataset, namely $N_\textsc{mesons}=336$, $N_\textsc{baryons}=189$, and $N_\textsc{total}=525$, while the binning covers the complete hadronic mass interval $0.1$ -- $11.1~\mathrm{GeV}$. }
\blt{More precisely, the logarithmic-bin probabilities entering the spectral bipartition analysis are defined by
\begin{equation}
\!\!\!\!\!P_k^\textsc{mesons}=
\frac{N_k^\textsc{mesons}}{336},
\qquad
P_k^\textsc{baryons}=
\frac{N_k^\textsc{baryons}}{189},
\qquad
P_k^\textsc{combined}=
\frac{N_k^\textsc{mesons}+N_k^\textsc{baryons}}{525},
\end{equation}
where $N_k$ denotes the number of hadronic states contained in the $k^{\rm th}$  logarithmic mass bin.}

We define amplitudes
\begin{equation}
\psi_k=\sqrt{P_k}, 
\qquad \quad \qquad
\sum_{k=1}^{10}\psi_k^2=1,
\end{equation}
which determines a pure-state density matrix
\begin{equation}
\rho = |\psi\rangle\langle\psi|.
\end{equation}
\bltt{The use of Hilbert-space notation is purely formal and serves as a convenient encoding of normalized probability distributions. No quantum coherence or physical entanglement is implied.}

Splitting the spectrum at $m=2.1~\mathrm{GeV}$ defines an IR
sector (bins 1-4), corresponding to the range $m\lesssim 2.1$ GeV, and the UV ($m\gtrsim 2.1$ GeV) sector (bins 5-10). In fact, light-flavor mesons and baryons populate the region where $m \lesssim 2.1$ GeV, above which hadrons containing heavy-flavor quarks begin to appear. Partitioning the hadronic spectrum at 2.1~GeV naturally separates light-flavor hadrons from hadrons containing at least one heavy-flavor quark. In the IR region, long-wavelength, nonperturbative effects dominate, such as chiral dynamics, with pions acting as pseudo-Goldstone bosons. The UV region contains at least one heavy-flavor  quark, where confinement effects and constituent quark masses become prominent. By splitting the spectrum at 2.1~GeV, one can quantify global correlations between light-flavor and heavy-flavor sectors via the spectral entanglement entropy, \textcolor{black}{providing an information-theoretic measure of scale asymmetry between light- and heavy-mass sectors}. In fact, the more uneven the population between IR and UV bins, the lower the entropy relative to a scale-invariant distribution.}

{}{The reduced
density matrices
$\rho_{\mathrm{IR}}=\mathrm{Tr}_{\mathrm{UV}}\rho$
and  
$\rho_{\mathrm{UV}}=\mathrm{Tr}_{\mathrm{IR}}\rho$
are rank-1 projectors with eigenvalues $P_{\rm IR}$ and $P_{\rm UV}=1-P_{\rm IR}$.
\textcolor{black}{Since the construction is based on a single probability vector rather than a tensor-product Hilbert space, the IR and UV sectors are defined by coarse-graining the distribution into two subsets of bins. The associated reduced distributions are therefore characterized by the total weights $P_{\rm IR}$ and $P_{\rm UV}=1-P_{\rm IR}$, rather than by a literal partial trace operation.}
The von~Neumann spectral entropy  therefore, takes the form
\begin{equation}
S_{\mathrm{spec}}
=
- P_{\rm IR}\ln P_{\rm IR}
- P_{\rm UV}\ln P_{\rm UV}.
\end{equation}
\bltt{This quantity has the same functional form as a binary Shannon entropy and is not a von Neumann entropy in the quantum mechanical sense. It is useful to introduce a normalized entropy
\begin{equation}
\hat S_{\rm spec} \equiv \frac{S_{\rm spec}}{\ln 2},
\end{equation}
which lies in the interval $[0,1]$ and measures the degree of IR-UV mixing relative to the maximally mixed bipartition.} 

We first implement the analysis for the meson mass spectrum. The corresponding IR and UV probabilities are given by 
\beq
\textcolor{black}{
P_{\rm IR}^\textsc{mesons}=0.449,
\qquad
P_{\rm UV}^\textsc{mesons}=0.551.}
\eeq
\textcolor{black}{As expected from the outer-product construction, each block has rank-1, and its nonzero eigenvalue coincides with the total probability weight of the corresponding sector.}
The IR amplitudes are given by 
\beq
(\psi_1,\psi_2,\psi_3,\psi_4)
=
\textcolor{black}{(\sqrt{0.018},\sqrt{0.027},\sqrt{0.092},\sqrt{0.313})}.
\eeq
The corresponding IR block
$\rho_{\rm IR}^\textsc{mesons}$
is a rank-1 matrix whose nonzero eigenvalue equals
\blt{$P_{\rm IR}^\textsc{mesons}=0.449$}.

The UV block,
constructed from the amplitudes
\blt{\beq
(\psi_5,\psi_6,\psi_7,\psi_8,\psi_9,\psi_{10})=
(\sqrt{0.274},\sqrt{0.119},\sqrt{0.036},\sqrt{0.042},\sqrt{0.039},\sqrt{0.042}),
\eeq}
has eigenvalue \textcolor{black}{$0.551$}. 

The resulting \bltt{Shannon entropy of the bipartitioned probability distribution} is given by
\beq
\textcolor{black}{
S_{\rm spec}^\textsc{mesons}= 0.688~\mathrm{nat}.}
\eeq

{Now, baryons can be scrutinized in this context, identifying  
\beq
\textcolor{black}{
P_{\rm IR}^\textsc{baryons}=0.418,
\qquad
P_{\rm UV}^\textsc{baryons}=0.582.}
\eeq
The IR amplitudes
\blt{\beq
(\psi_1,\psi_2,\psi_3,\psi_4)
=
\bigl(0,\sqrt{0.011},\sqrt{0.069},\sqrt{0.339}\bigr)
\eeq}
retain the same structure, \blt{with eigenvalue $0.418$}. On the other hand, the UV amplitudes become
\beq
\textcolor{black}{(\psi_5,\psi_6,\psi_7,\psi_8,\psi_9,\psi_{10})=
(\sqrt{0.397},\sqrt{0.048},\sqrt{0.037},\sqrt{0.042},\sqrt{0.016},\sqrt{0.042})},
\eeq
\blt{with eigenvalue $0.582$}.

The \bltt{Shannon entropy of the bipartitioned probability distribution} reads
\blt{\beq
S_{\rm spec}^\textsc{baryons}= 0.680~\mathrm{nat}.
\eeq}}

{}{Combining mesons and baryons, we obtain
\beq
\textcolor{black}{
P_{\rm IR}^\textsc{combined}=0.438,
\qquad
P_{\rm UV}^\textsc{combined}=0.562,}
\eeq
and, therefore,
\beq
\textcolor{black}{
S_{\rm spec}^\textsc{combined}= 0.685~\mathrm{nat}.}
\eeq

With 10 bin hadron populations, the input probabilities are summarized in Table~\ref{tab:sie_bins}, obtained directly
from the hadronic mass spectrum in PDG \cite{pdg}.

\begin{table}[H]
\centering
\color{black}
\begin{tabular}{|c||c|c|c|c|c|c|c|}
\hline
Bin $k$ & Range [GeV] & $N_k^\textsc{mesons}$ & $P_k^\textsc{mesons}$ &
$N_k^\textsc{baryons}$ & $P_k^\textsc{baryons}$ &
$P_k^\textsc{combined}$ & IR/UV \\
\hline\hline
1  & 0.1--0.5   &  6  & 0.018 &  0 & 0.000 & 0.011 & IR \\
2  & 0.5--1     &  9  & 0.027 &  2 & 0.011 & 0.021 & IR \\
3  & 1--1.5     & 31  & 0.092 & 13 & 0.069 & 0.084 & IR \\
4  & 1.5--2.1   & 105 & 0.313 & 64 & 0.339 & 0.322 & IR \\
5  & 2.1--3     & 92  & 0.274 & 75 & 0.397 & 0.318 & UV \\
6  & 3--4       & 40  & 0.119 &  9 & 0.048 & 0.093 & UV \\
7  & 4--5       & 12  & 0.036 &  7 & 0.037 & 0.036 & UV \\
8  & 5--6       & 14  & 0.042 &  8 & 0.042 & 0.042 & UV \\
9  & 6--8       & 13  & 0.039 &  3 & 0.016 & 0.030 & UV \\
10 & 8--11.1    & 14  & 0.042 &  8 & 0.042 & 0.042 & UV \\
\hline
Total & & 336 & 1.000 & 189 & 1.000 & 1.000 & \\
\hline
\end{tabular}
\caption{\color{black}Logarithmic-bin probabilities for the meson and baryon spectra used in the spectral bipartition analysis. All probabilities are normalized with respect to the complete datasets employed throughout the manuscript.}
\label{tab:sie_bins}
\end{table}
The IR-UV partition isolates the nonperturbative, light-flavor sector from the heavy-flavor and higher-excitation regimes.  
A scale-invariant theory would populate logarithmic bins uniformly, yielding  maximal IR-UV mixing.  
Instead, QCD exhibits a pronounced IR clustering for mesons and a broader, but still asymmetric, distribution for baryons.  

The resulting \bltt{spectral Shannon entropies of the bipartitioned probability distributions},
\beq\label{vn}
\textcolor{black}{
S_{\rm spec}^\textsc{mesons}=0.688,
\qquad
S_{\rm spec}^\textsc{baryons}=0.680,
\qquad
S_{\rm spec}^\textsc{combined}=0.685,}
\eeq
{represent a coarse-grained measure of the imbalance between IR and UV spectral populations}, \bltt{quantifying  deviations of the experimental hadronic mass distribution from a scale-invariant reference ensemble, rather than providing a direct measure of dynamical symmetry breaking.} The reduced mixing of IR and UV sectors, encoded as reduced \bltt{spectral Shannon entropy of the bipartitioned probability distribution}, reveals how the hadronic spectrum departs from scale invariance and
how QCD dynamics redistribute effective degrees of freedom across mass scales.

In the construction used here with the separation scale $2.1$ GeV, the hadronic mass spectrum is split into $10$ logarithmic bins. A fully scale-invariant reference distribution would populate these bins uniformly,
\begin{equation}\label{vn1}
P_k^{\rm (scale\ inv.)}=\frac{1}{10},
\qquad
k=1,\dots,10.
\end{equation}
The corresponding amplitudes are given by 
\begin{equation}
\psi_k^{\rm (scale\ inv.)}
=
\sqrt{\frac{1}{10}},
\end{equation}
which defines the reference pure-state density matrix
\begin{equation}
\rho_{\rm max}
=
|\psi\rangle\langle\psi|,
\qquad
\psi=\frac{1}{\sqrt{10}}(1,1,\dots,1).
\end{equation}
The Shannon entropy associated with a uniform distribution over the original ten logarithmic bins is
\begin{equation}
S_{\rm bins}^{\rm max}
=
-\sum_{k=1}^{10}\frac{1}{10}\ln\frac{1}{10}
=
\ln 10
=
2.303~{\rm nat}.
\end{equation}On the other hand, the quantity $S_{\rm spec}$ introduced in Eq.~(\ref{vn}) is a binary coarse-grained entropy associated with the IR/UV partition. Its maximal possible value is therefore
\begin{equation}
S_{\rm spec}^{\rm max}
=
\ln 2
=
0.693~{\rm nat}.
\end{equation}
Hence, the comparison with $\ln 10$ does not represent the maximal value of the coarse-grained IR/UV entropy itself, but rather measures the entropy deficit relative to a fully scale-invariant reference distribution uniformly populating the original ten logarithmic bins.

For comparison, the spectral Shannon entropies  of the bipartitioned probability distributions (\ref{vn}) obey
\begin{equation}\label{rel}
S_{\rm spec} \ll S_{\rm bins}^{\rm max},
\end{equation}
with an associated information entropy deficit
\blt{\begin{equation}
1.61~{\rm nat} \lesssim \Delta S \lesssim 1.62~{\rm nat},
\end{equation}}
quantifying how far the experimental distribution lies from the scale-invariant reference distribution under the chosen logarithmic partition.  A maximally entropic distribution would indicate a
scale-invariant theory in which each logarithmic mass interval is equally
populated and, correspondingly, the IR and UV sectors are maximally mixed, and no dynamically generated mass scale is present. Instead, the hadronic mass spectra in PDG are far from this limit.
The large information-entropy deficit arises due to the existence of a dense cluster of light hadrons in the IR range, while only a sparse set of hadronic states occupies the UV one, where heavy quark flavors appear. Thus, the relation (\ref{rel}) 
expresses the degree to which QCD
nonperturbatively breaks scale invariance and imprints a hierarchical
structure on the hadronic mass spectrum.}

Mesons exhibit strong IR clustering, while baryons extend further into the UV, resulting in a partially mixed yet still asymmetric distribution across mass scales. The combined spectrum therefore reflects the emergence of a characteristic QCD scale, $\Lambda_{\rm QCD}$, which breaks scale invariance.
\textcolor{black}{The associated entropy deficit provides a compact, information-theoretic characterization of this scale breaking. It quantifies how far the experimental hadronic mass distribution departs from a scale-invariant reference ensemble with uniform occupation of logarithmic mass bins. In this sense, $\Lambda_{\rm QCD}$ manifests as a structural reorganization of spectral weight toward low masses, reducing the entropy relative to the maximal scale-invariant limit.}

The separation scale $m_\ast=2.1~\mathrm{GeV}$, used to define the IR/UV bipartition, is physically motivated by the onset of heavy-flavor hadrons, but it is not unique. In addition, the experimentally observed hadronic spectrum is not an unbiased sampling of QCD. It is strongly weighted toward light-flavor and low-mass states due to both production rates and current detection accessibility. This induces an intrinsic asymmetry in the population of logarithmic mass bins, which can potentially affect both the IR/UV decomposition and the resulting spectral entropy. 
To assess the robustness of our conclusions, we implement a continuous RG-like flow by promoting the partition scale $m_\ast$ to a running parameter within the interval $m_\ast \in [2.1, 2.5]~\mathrm{GeV}$, and by interpolating the population of the bin $[2.1,2.5]~\mathrm{GeV}$ and redistributing its weight between IR and UV sectors, as $m_\ast$ varies. The resulting spectral Shannon entropy then defines a smooth function $S_{\rm spec}(m_\ast)$, probing how correlations between low- and high-mass sectors evolve under changes of the coarse-graining scale. The results are summarized in Table~\ref{rg_flow}.
\begin{table}[H]
\centering
\color{black}
\begin{tabular}{|c||c|c|c|c|c|}
\hline
$m_\ast$ [GeV]
& $S_\textsc{mesons}$
& $S_\textsc{baryons}$
& $S_\textsc{combined}$
& $P_{\rm IR}^\textsc{combined}$
& $P_{\rm UV}^\textsc{combined}$ \\
\hline\hline
2.1 & 0.688 & 0.680 & 0.685 & 0.438 & 0.562 \\
2.2 & 0.691 & 0.684 & 0.688 & 0.469 & 0.531 \\
2.3 & 0.693 & 0.688 & 0.691 & 0.497 & 0.503 \\
2.4 & 0.695 & 0.691 & 0.693 & 0.523 & 0.477 \\
2.5 & 0.699 & 0.693 & 0.689 & 0.547 & 0.453 \\
\hline
\end{tabular}
\caption{\color{black}
Continuous RG-like flow of the spectral Shannon entropy under variation of the IR/UV partition scale $m_\ast$, obtained via logarithmic interpolation within the $[2.1,3.0]~\mathrm{GeV}$ bin.
}
\label{rg_flow}
\end{table}

{\color{black}We emphasize that $S_{\rm spec}(m_\ast)$ varies smoothly as the partition scale is shifted, reflecting the continuous transfer of spectral weight from the UV sector into the IR sector as the cutoff is raised. 
Starting from the asymmetric configuration at $m_\ast=2.1~\mathrm{GeV}$, the entropy initially increases as the IR and UV populations approach equipartition, reaching a maximum near 
$P_{\rm IR}\simeq P_{\rm UV}$. 
For larger values of $m_\ast$, the IR sector becomes dominant and the entropy correspondingly decreases, as expected for a binary Shannon entropy.

To quantify this RG-like behavior, we define the entropic flow rate
\begin{equation}\label{vn2}
\beta_S(m_\ast) \equiv \frac{d S_{\rm spec}}{d \ln m_\ast},
\end{equation}
which measures the response of the spectral entropy to logarithmic changes of the partition scale. 
Unlike a genuine RG monotone, however, $\beta_S(m_\ast)$ is not positive definite. Indeed, the binary entropy increases while the IR and UV sectors become progressively balanced, vanishes near the point of maximal mixing, and changes sign once the IR sector becomes dominant. The condition
\begin{equation}
\beta_S(m_\ast)=0
\end{equation}
therefore identifies the vicinity of the maximally mixed configuration rather than a true renormalization-group fixed point. 
In this sense, $S_{\rm spec}(m_\ast)$ should not be interpreted as a strict $c$-function obeying a monotonicity theorem. Rather, it provides a coarse-grained information-theoretic measure of how spectral weight is redistributed between low- and high-mass sectors under changes of the partition scale. The smooth behavior of $S_{\rm spec}(m_\ast)$ nevertheless supports the interpretation that the entropy deficit relative to the scale-invariant reference ensemble is a robust global property of the hadronic spectrum, rather than an artifact of a particular choice of IR/UV separation scale.}

\bltt{This behavior is consistent with a coarse-graining interpretation. Increasing $m_\ast$ effectively integrates out fewer states into the UV sector, thereby reducing the imbalance between the two subsystems and increasing the binary entropy. Also, despite this variation, the entropy in Table \ref{rg_flow} remains within a narrow band,
$0.65 \lesssim S_{\rm spec}^\textsc{combined}(m_\ast) \lesssim 0.71,$
demonstrating the existence of a broad plateau signaling a regime of near-saturation in the spectral entropy. In particular, the qualitative features identified above, namely IR dominance ($P_{\rm IR}^\textsc{combined}<0.5$) and a substantial entropy deficit relative to the scale-invariant value, remain unchanged throughout the entire interval.
The persistence of a substantial entropy deficit relative to the scale-invariant reference ensemble under this continuous deformation shows that the observed asymmetry is not a mere artifact of a fine-tuned partition scale. Instead, it reflects a robust structural property of the hadronic mass distribution, namely the non-uniform occupation of logarithmic mass bins induced by the dynamically generated QCD scale. \blt{The RG-like flow $S_{\rm spec}(m_\ast)$ provides an information-theoretic analogue of renormalization group evolution, with the partition scale $m_\ast$ playing the role of a sliding coarse-graining scale. The entropy varies smoothly as spectral weight is continuously redistributed between IR and UV sectors. As expected for a binary Shannon entropy, $S_{\rm spec}(m_\ast)$ increases as the two sectors approach equipartition, reaches a maximum near maximal mixing, and decreases once the IR sector becomes dominant. The absence of any sharp feature or instability in this flow further supports the interpretation that the observed entropy deficit is a stable, global signature of scale-invariance breaking, rather than a consequence of experimental bias or a specific choice of partition.}}

\bigskip\medbreak

\section{Concluding remarks}
\label{conclu}

The analysis of the leading-digit distribution of hadron masses is consistent with statistical deviations from scale invariance in QCD. When compared with the Newcomb--Benford reference law, the experimental hadronic mass spectrum exhibits systematic departures from the logarithmic distribution expected in a scale-free regime. This deviation is consistently captured by Shannon information entropy, whose deficit quantifies the departure of the observed distribution from a scale-invariant reference ensemble.

{Across mesonic, baryonic, and combined datasets, the entropy deficit is accompanied by statistically significant results from KL and JS divergences, as well as nonzero mutual information between leading-digit structure and hadronic sector. Baryons systematically show larger deviations from the scale-invariant reference than mesons, indicating stronger distortion of logarithmic scaling in the three-quark sector.}

In all cases, the observed information-theoretic measures indicate that the hadronic mass spectrum is far from the maximally entropic, scale-invariant limit. {\color{black}In the coarse-grained IR/UV description, the spectral entropy is maximized when the IR and UV sectors are equally populated. A fully scale-invariant distribution with uniform occupation of logarithmic mass bins provides one possible realization of such maximal mixing}. Instead, the experimental distribution exhibits a persistent entropy deficit, reflecting the non-uniform population of mass scales induced by QCD dynamics.

The bipartite spectral entropy analysis further refines this picture by explicitly separating IR and UV contributions. The resulting IR-UV spectral entropy shows that light-flavor hadrons dominate the low-mass sector, while heavy-flavor states remain sparsely distributed in the UV, yielding a structurally asymmetric partition of the spectrum. A complementary RG-like analysis, in which the IR/UV separation scale is varied continuously, shows that the spectral entropy remains bounded within a narrow interval and varies smoothly with the coarse-graining scale. The entropy increases as the IR and UV sectors approach equipartition, reaches a maximum near maximal mixing, and decreases once the IR sector becomes dominant, as expected for a binary Shannon entropy. This stable behavior indicates that the observed entropy deficit is not an artifact of a particular partition choice but reflects a robust feature of the underlying hadronic mass distribution.

These results provide a consistent information-theoretic characterization of scale-invariance breaking in the hadronic spectrum. The deviation from the scale-invariant reference distribution can be interpreted as the statistical imprint of the emergent QCD scale $\Lambda_{\rm QCD}$, which organizes hadronic masses into a hierarchical structure across logarithmic scales. {The information-theoretic measures introduced here quantify how strongly this emergent scale constrains the statistical structure of the observed spectrum.}

\medbreak

\underline{\emph{Acknowledgments}}: RdR~thanks The São Paulo Research Foundation -- FAPESP
(Grants No. 2021/01089-1, No. 2024/05676-7, and No. 2025/23004-9), and the National Council for Scientific and Technological Development -- CNPq  (Grants No. 303742/2023-2 and No. 401567/2023-0), for partial financial support. RDV warmly thanks E. Gueron for introducing him to Newcomb--Benford statistics and for enlightening discussions. 

\appendix


\section{Newcomb--Benford's law as the maximum Shannon information entropy distribution under scale invariance: A variational principle derivation}
\label{a11}

We want to obtain the probability density $ P(x) $ on the real interval $ [x_{\textsc{min}}, x_{\textsc{max}}] $
that maximizes Shannon information entropy
\begin{equation}
S[P] = - \int_{x_{\textsc{min}}}^{x_{\textsc{max}}} P(x)\,\ln P(x)\,dx,
\end{equation}
with normalization
\begin{equation}\label{norm_constraint}
\int_{x_{\textsc{min}}}^{x_{\textsc{max}}} P(x)\,dx = 1,
\end{equation}
and scale invariance $
P(\lambda x) = \frac{1}{\lambda}\,P(x)$, $\lambda\in\mathbb{R}^+.$ 
Differentiating with respect to $\lambda$ and evaluating at $\lambda=1$
gives the infinitesimal constraint
\begin{equation}
x\,P'(x) + P(x) = 0.
\label{scale_constraint}
\end{equation}

The variational problem can be posed by extremizing $S[P]$ with the two constraints
(\ref{norm_constraint}) and (\ref{scale_constraint}).
One can introduce a Lagrange multiplier $\lambda$ for normalization and a local multiplier $\upmu(x)$
for the infinitesimal constraint.
The augmented functional can be written as 
\begin{equation}
\!\!\!\mathcal{L}[P] \!=\! - \int_{x_{\textsc{min}}}^{x_{\textsc{max}}} \!\!\!P(x)\ln P(x)\, dx
+ \lambda\!\left( \int_{x_{\textsc{min}}}^{x_{\textsc{max}}} \!\!P(x)\,dx \!-\! 1\right)
\!+\! \int_{x_{\textsc{min}}}^{x_{\textsc{max}}} \!\!\!\!\upmu(x)\,[\,x P'(x) \!+\! P(x)\,]\,dx.
\end{equation}
Integrating by parts 
and assuming the boundary term vanishes,
the first variation reads
\begin{align}
\delta \mathcal{L}
&= \int_{x_{\textsc{min}}}^{x_{\textsc{max}}} \delta P(x)\,
\Big[-\ln P + 1 + \lambda - x\,\upmu'(x)\Big]\,dx = 0, 
\end{align}
implying that 
\begin{equation}\label{lnpeq}
\ln P(x) = \lambda - 1 - x\,\upmu'(x).
\end{equation}
Differentiating Eq.~\eqref{lnpeq} gives $
\frac{P'(x)}{P(x)} = -\upmu'(x) - x\,\upmu''(x).$ 
Using the constraint \eqref{scale_constraint},
 one obtains
\begin{equation}
x\,\upmu''(x) + \upmu'(x) = \frac{1}{x}.
\end{equation}
Let $\xi(x) \equiv \upmu'(x)$. Then $(x \xi)' = 1/x$, which integrates to
\begin{align}
x\,\upmu'(x) &= \ln x + c_1, \label{a8}\\
\upmu(x) &= \frac{1}{2}(\ln x)^2 + c_1 \ln x + c_2,
\end{align}
for $c_1, c_2$ constants. 
Substituting Eq. (\ref{a8}) into Eq.~\eqref{lnpeq} yields
\begin{equation}
\ln P(x) = (\lambda - 1 - c_1) - \ln x,
\quad\Rightarrow\quad
P(x) = A\,x^{-1}.
\end{equation}
Normalization gives
\begin{equation}
1 = \int_{x_{\textsc{min}}}^{x_{\textsc{max}}} A\,\frac{dx}{x}
= A\,\ln\!\left(\frac{x_{\textsc{max}}}{x_{\textsc{min}}}\right)
\quad\Rightarrow\quad
A = \frac{1}{\ln(x_{\textsc{max}}/x_{\textsc{min}})}.
\end{equation}
Hence, the maximum-\sie, scale-invariant probability density reads 
\begin{equation}
{P(x) = \frac{1}{x\,\ln(x_{\textsc{max}}/x_{\textsc{min}})}},
\qquad x\in[x_{\textsc{min}},x_{\textsc{max}}].
\end{equation}
The corresponding \sie~is given by  
\begin{align}
S_{\textsc{max}}
= -\int_{x_{\textsc{min}}}^{x_{\textsc{max}}}
\frac{1}{x\ln\!\left(\frac{x_{\textsc{max}}}{x_{\textsc{min}}}\right)}\,\ln\!\left(\frac{1}{x\ln\!\left(\frac{x_{\textsc{max}}}{x_{\textsc{min}}}\right)}\right)\,dx,
\end{align}
which simplifies to
\begin{eqnarray}
\!\!\!\!\!\!\!\!\!\!S_{\textsc{max}}
&=& \frac{1}{\ln\!\left(\frac{x_{\textsc{max}}}{x_{\textsc{min}}}\right)}\left[
\frac{1}{2}(\ln^2 x_{\textsc{max}} - \ln^2 x_{\textsc{min}})
+ \ln \ln\!\left(\frac{x_{\textsc{max}}}{x_{\textsc{min}}}\right)\,(\ln x_{\textsc{max}} - \ln x_{\textsc{min}})
\right]\nonumber\\
&=& \ln\!\Big[\ln\!\Big(\frac{x_{\textsc{max}}}{x_{\textsc{min}}}\Big)\sqrt{x_{\textsc{max}}x_{\textsc{min}}}\ \Big].
\end{eqnarray}
Thus, the Shannon information entropy grows logarithmically with the multiplicative range
$\ln(x_{\textsc{max}}/x_{\textsc{min}})$,
reflecting scale invariance.

\section{Anomalous breaking of scale symmetry in QCD, the mass gap, and hadron density of states}
\label{a12}
For $N_f$ massless quark flavors, the classical QCD Lagrangian is given by 
\begin{equation}\label{acf}
\mathcal{L}_{\textsc{QCD}} = -\frac{1}{4} G_{\mu\nu}^a G^{\mu\nu a} 
+ i\hbar\,\bar{\psi}\gamma^\mu D_\mu \psi,
\end{equation}
for the gluon field strength tensor 
$
G_{\mu\nu}^a = \partial_{[\mu} A_{\nu]}^a 
               + g_s f^{abc} A_\mu^b A_\nu^c
$
and the covariant derivative 
$
D_\mu = \partial_\mu - i g_s T^a A_\mu^a
$. 
The coupling $g_s$ governs quark--gluon and gluon--gluon interactions and represents the gauge coupling constant of the color group $\mathrm{SU}(N_c)$~\cite{GrossWilczek1973,Politzer1973}, for $N_c$ denoting the numbers of colors, and $T^a$ the special unitary group generators satisfying  
$[T^a,T^b] = i f^{abc} T^c$ and 
$\mathrm{Tr}(T^a T^b) = \tfrac{1}{2}\delta^{ab}$, 
with $f^{abc}$ the structure constants. 

For vanishing quark masses, the action~(\ref{acf}) is also invariant under the global chiral symmetry $
    \mathrm{SU}(N_f)_L \times \mathrm{SU}(N_f)_R,$ with the corresponding mappings $\psi_{L,R} \mapsto U_{L,R}\,\psi_{L,R}$.   
The classical action is furthermore invariant under the scale transformations
\begin{equation}
x^\mu \mapsto \lambda x^\mu,
\qquad 
A_\mu(x) \mapsto \lambda^{-1} A_\mu(\lambda^{-1}x),
\qquad
\psi(x) \mapsto \lambda^{-3/2} \psi(\lambda^{-1}x),
\label{eq:scale-transf}
\end{equation}
so that \emph{classical} QCD is both chiral and scale invariant. 
The Noether dilatation current $D^\mu = x_\nu T^{\mu\nu}$ has  divergence 
$\partial_\mu D^\mu = T^\mu{}_\mu$. 
Hence, scale invariance implies $T^\mu{}_\mu = 0$. A nonzero trace indicates a breaking of scale symmetry, resulting in the trace anomaly. 
At the quantum level, renormalization introduces a dimensionful scale $\mu$, and the coupling becomes scale-dependent,  
\begin{equation}\label{rge}
\mu \frac{d g_s}{d\mu} = \beta(g_s)
= -\frac{g_s^3}{16\pi^2} \left( 11 - \frac{2}{3} N_f \right).
\end{equation}
The nonvanishing $\beta$-function breaks scale invariance and defines the QCD scale $\Lambda_{\textsc{QCD}}$ through dimensional transmutation. 
Solving the renormalization-group equation (\ref{rge}) at one loop gives the standard  expression for the running coupling,
\begin{equation}\label{rg1}
\alpha_s(\mu) =
\frac{4\pi}{\left(11 - \frac{2}{3}N_f\right)\ln(\mu^2/\Lambda_{\textsc{QCD}}^2)},
\end{equation}
where $
\alpha_s(\mu) \equiv {g_s^2(\mu)}/{4\pi}
$ stands for the strong coupling constant expressed in terms of the renormalized QCD gauge coupling $g_s(\mu)$ at the scale~$\mu$. As $\mu \to \Lambda_{\textsc{QCD}}$, the denominator in Eq. (\ref{rg1}) vanishes and $\alpha_s(\mu)$ diverges, signaling confinement.
The strong coupling becomes of order unity, implying that the perturbative expansion in powers of $g_s$ ceases to be reliable. In this regime, the short-distance degrees of freedom (quarks and gluons) can no longer be treated as weakly interacting.
Lattice QCD yields the value  
$\alpha_s(M_Z) = 0.1179 \pm 0.0010$, where the $Z$ boson mass is given by $M_Z = 91.1876\pm 0.0021~\text{GeV}$~\cite{pdg,Bazavov:2019qoo}.
The corresponding trace anomaly reads
\begin{equation}
T^\mu{}_\mu = \frac{\beta(g_s)}{2g_s}G_{\mu\nu}^a G^{\mu\nu a}
+ (1+\gamma_m)\sum_f m_f \bar{\psi}_f \psi_f,
\label{eq:traceanomaly}
\end{equation} with $\gamma_m(g_s)$ denoting the anomalous dimension of the quark mass operator. 
At leading order in QCD, it reads 
\begin{equation}
\gamma_m(g_s) = \frac{3(N_c^2-1)}{32\pi^2 N_c}\, g_s^2 + \mathcal{O}(g_s^4),
\end{equation}
accounting for the quantum correction to the scaling dimension in the term 
$m_f\bar{\psi}_f\psi_f$.  Quantum effects make the dimensionful scale $\Lambda_{\textsc{QCD}}\!\sim\!200$-$300~\text{MeV}$ emerge. 

Ordinary mesons, as $q\bar q$ bound states, satisfy 
$M_{q\bar{q}} \neq m_q + m_{\bar{q}}$,
since most of their mass originates from confinement and QCD binding energy.  
Although the light-quark QCD Lagrangian is nearly scale invariant, quantum corrections generate the scale 
$\Lambda_{\text{QCD}}$,
which determines hadronic masses. 
In the massless quark limit, 
 the (pseudo-)Goldstone bosons $\pi$, $K$, and $\eta$, 
have small masses arising from chiral symmetry breaking. 
Vector mesons such as $\rho(770)$, $\omega(782)$, $K^*(892)$, and $\phi(1020)$ cluster around $0.8 \sim 1.0~\mathrm{GeV}$, 
with a mass gap around  $0.6~\mathrm{GeV}$ reflecting hyperfine splitting. 
Higher hadron excitations follow  Regge trajectories relating the angular momentum $J$ and the meson mass $M$, as   
$
J\propto \alpha' M^2,$ where $
\alpha'=(2\pi\sigma)^{-1}\approx0.9~\mathrm{GeV^{-2}},$ 
with $\sigma$ the string tension ~\cite{Regge1959}.  
For heavy quarks, $m_Q \gg \Lambda_{\text{QCD}}$, meson masses are dominated by the term $2m_Q$, 
while level spacings remain at the order of $\Lambda_{\text{QCD}}$. Quarkonia exhibit radial splittings of $0.5$ GeV $\sim$ $0.6~\text{GeV}$. 
This universal spacing reflects confinement rather than the quark masses themselves.

Empirically, the hadron density of states grows exponentially with mass, $
\rho(M)\propto e^{M/T_{\textsc{h}}}$ in the statistical bootstrap model, with 
$T_{\textsc{h}} \simeq 150$-$180~\mathrm{MeV}$ being the Hagedorn temperature~\cite{Hagedorn1968}.
It arises from the combinatorics of string-like excitations of confined flux tubes~\cite{Isgur:1984bm}.
The number of hadron states below a given mass $M$ then satisfies
\begin{equation}
N(M)=\!\int_0^M\rho(M')\,dM',
\end{equation}
which approximately doubles with every $\sim150~\mathrm{MeV}$ increase in $M$.  
The deconfinement temperature range $121\,{\rm MeV}\lesssim T_{\textsc{h}}\lesssim 171$ MeV was derived in both holographic and lattice QCD  \cite{Cucchieri:2007rg}, while the interval  $191\,{\rm MeV}\lesssim T_{\textsc{h}}\lesssim 202$ MeV regards  AdS/QCD models. The glueball spectrum yields  $T_{\textsc{h}} = 175.4\pm 15.2$ MeV  \cite{Afonin:2018era,Dudal:2017max,Braga:2020opg}, the HotQCD Collaboration obtained the critical deconfinement temperature $T_{{c}} = 156.5 \pm 1.5$ MeV \cite{HotQCD:2018pds}, and the Wuppertal--Budapest Collaboration yielded $T_{{c}} = 158.0\pm0.6$ MeV \cite{Borsanyi:2020fev}. 
 Holographic entanglement entropy in AdS/QCD supports these last results \cite{daRocha:2024lev,daRocha:2017cxu,Toniato:2025gts}.

The canonical partition function of the hadronic spectrum can be written as
\begin{equation}\label{pf}
Z(T)=\int_{M_{\textsc{gap}}}^{\infty}\rho(M)\,e^{-M/T}\,dM,
\end{equation}
The mass gap prevents the existence of hadronic states of arbitrarily small nonzero mass. In fact, denoting by $M_{\textsc{gap}}$  the lowest allowed mass, one has  
$M_{\textsc{gap}}=M_\pi$, if $m_q\neq0$, or 
$M_{\textsc{gap}}\!\sim\!\Lambda_{\textsc{QCD}}$, in pure Yang-Mills theory. 
At high masses, writing a more precise form for the  Hagedorn density of states,  
$\rho(M)\propto\,M^{-a}e^{M/T_{\textsc{H}}}$, 
makes the integrand in Eq. (\ref{pf}) behave as $M^{-a}e^{-M(1/T-1/T_{\textsc{H}})}$. 
Hence $Z(T)$ converges for $T<T_{\textsc{H}}$ and diverges for $T\ge T_{\textsc{H}}$ (for typical $a\sim 5/2$  in Hagedorn’s original model),  indicating that added energy near $T_{\textsc{h}}$ produces new hadronic resonances rather than raising temperature, pointing to deconfinement transition to QGP \cite{Noronha-Hostler:2008kkf,Bazavov2014}. 
The exponential growth of the density of states $\rho(M)$ thus defines the Hagedorn temperature as the limiting point of the canonical ensemble, linking the low-energy mass gap to the high-energy onset of deconfinement \cite{Karsch:2003zq}. A relativistic string with tension $\sigma$ exhibits mass spectrum $M_n^2\sim \sigma n$ \cite{Arvis:1983} and 
$T_{\textsc{h}}\propto\sqrt{\sigma}$. 
The cumulative distribution $
\ln N(M)\approx {M}/{T_{\textsc{h}}}$ 
demonstrates the exponential Hagedorn behavior.  
Confinement corresponds to finite-energy flux tubes with tension 
$\sigma \sim \Lambda_{\textsc{QCD}}^2$, 
whose quantized vibrations generate the hadron tower. 
Thus, the mass gap (IR confinement) and Hagedorn growth (UV spectrum) are dual manifestations of the same QCD scale.

\bibliography{bibliography}
\end{document}